\RequirePackage{rotating}
\documentclass[twocolumn]{svjour3}          

\usepackage{graphicx}

\usepackage{url}

\usepackage{booktabs}
\usepackage{longtable}
\usepackage{tabularx}
\usepackage{pdflscape}
\usepackage{appendix}
\usepackage{tikz}

\begin{document}

\title{Modeling interdependent privacy threats}


\author{Shuaishuai Liu 
        \and
        Gergely Bicz{\'o}k}

\institute{CrySyS Lab
Budapest Univ. of Technology and Economics, Budapest, Hungary\\
          \email{\{sliu,biczok\}@crysys.hu}
}
\date{Received: date / Accepted: date}

\maketitle

\begin{abstract}
The rise of online social networks, user-gene-rated content, and third-party apps made data sharing an inevitable trend, driven by both user behavior and the commercial value of personal information. As service providers amass vast amounts of data, safeguarding individual privacy has become increasingly challenging. Privacy threat modeling has emerged as a critical tool for identifying and mitigating risks, with methodologies such as LINDDUN, xCOMPASS, and PANOPTIC offering systematic approaches. However, these frameworks primarily focus on threats arising from interactions between a single user and system components, often overlooking interdependent privacy (IDP); the phenomenon where one user’s actions affect the privacy of other users and even non-users. IDP risks are particularly pronounced in third-party applications, where platform permissions, APIs, and user behavior can lead to unintended and unconsented data sharing, such as in the Cambridge Analytica case.

We argue that existing threat modeling approaches are limited in exposing IDP-related threats, potentially underestimating privacy risks. To bridge this gap, we propose a specialized methodology that explicitly focuses on interdependent privacy. Our contributions are threefold: (i) we identify IDP-specific challenges and limitations in current threat modeling frameworks, (ii) we create IDPA, a threat modeling approach tailored to IDP threats, and (iii) we validate our approach through a case study on WeChat. We believe that IDPA can operate effectively on systems other than third-party apps and may motivate further research on specialized threat modeling.
\end{abstract}

\section{Introduction}
\label{sec:intro}
Since the advent of online social networks, social apps, and the shift toward user-generated content, data sharing has become an irrevocable trend. On top of user preferences and general pro-sharing behavior, personal information also holds immense commercial value~\cite{acquisti2016economics}. 
The promise of vast profits from the big data ecosystem and its ever-hungry machine learning models strongly incentivizes service providers to hoard massive amounts of personal data. Under these circumstances, protecting the privacy of individuals is becoming increasingly challenging.

To counter these challenges, privacy threat modeling (PTM) has emerged as a crucial first step in the quest for privacy protection. If done properly, threat modeling enables risk identification, assessment, and treatment, as well as regulatory compliance and the formulation of privacy requirements for privacy engineering. To this end, multiple systematic PTM approaches have emerged over time, e.g., LINDDUN~\cite{wuyts2015linddun}, and lately, xCOMPASS~\cite{maps_xcompass} and PANOPTIC~\cite{katcherpanoptic}.
Meanwhile, as online services and personas have become more and more intertwined, the phenomenon of interdependent privacy (IDP) was born~\cite{DBLP:conf/fc/BiczokC13}.  IDP denotes the situation where the privacy-related actions of others can influence an individual's privacy. Although generally under-studied, a domain that has received significant attention regarding IDP is third-party apps; see, e.g., the 2019 Cambridge Analytica debacle~\cite{DBLP:journals/compsec/SymeonidisBSPSP18}. In this domain, IDP risks arise owing to the combination of existing platform permissions and API features and the actions of fellow users, potentially resulting in personal data sharing without the knowledge and consent of the affected individuals.

We argue that the inherent nature of IDP-related threats makes them hard to identify using elaborate but generic PTM approaches such as LINDDUN. Apart from its documented issues with usability, a key limitation of LINDDUN (and other existing PTM techniques) is that it focuses on the direct privacy threats between a generic user and the system components. By treating the threats stemming from the interplay of multiple users and their data flows \emph{implicitly}, traditional PTM approaches potentially miss out on IDP situations and, therefore, underestimate privacy risks such as sharing others' data (e.g., simply sharing your address book with a mobile app~\cite{liu2025idpfilter}) and inference based on data correlation (e.g., kin genomic privacy~\cite{DBLP:conf/ccs/HumbertAHT13}).

We advocate for a specialized PTM methodology to effectively identify IDP-related threats: the \emph{explicit} focus on such threats empowers threat modelers (potentially even privacy practitioners not trained in IDP) not to ignore privacy risks relevant to the system under study. Despite its limitations, LINDDUN could serve as a basis for customization owing to its focus on data flows. Note that other PTM frameworks built on different concepts, such as xCOMPASS (based on attacker personas) or PANOPTIC (based on privacy context), might also be potential starting points (although the customization process would likely be different).

In this paper, we study threat modeling with a keen focus on IDP-related privacy threats. Specifically, our contribution is three-fold. First, we identify IDP-specific challenges and tool-specific limitations in terms of privacy threat modeling and define principles for improvement. These principles are inspired by the 3R framework~\cite{kamleitner2019your} and our own 6A principles, and manifest in classifying data flows, defining specialized threat categories, and differentiating among misactors. Second, we create a methodology along these steps to effectively find IDP-related threats. Third, we present a case study on the WeChat mini-app platform to validate our methodology and demonstrate its utility in a real-world scenario. Although we use the low-hanging fruit of third-party apps as both motivation and case study, the presented methodology generalizes to arbitrary (software) systems.

The rest of this paper is organized as follows. Section~\ref{sec:bg} introduces the necessary background concerning IDP, including real-world incidents, and privacy threat modeling. Section~\ref{sec: shortcomings} summarizes the unique challenges faced in mitigating IDP issues and analyzes the IDP-specific shortcomings of state-of-the-art privacy threat modeling techniques. Section~\ref{sec:principles} lays out the design principles of an IDP-specific threat modeling approach covering threat elicitation, analysis and mitigation. Next, Section~\ref{sec:ida} defines IDPA, a customized threat modeling approach specifically tailored to IDP-related threats. Section~\ref{sec:case} presents our WeChat case study, showcasing the effectiveness and operability of IDPA. Finally, Section~\ref{sec:conc} concludes the paper.

\section{Background}
\label{sec:bg}
In this section, we briefly introduce the necessary background for this study regarding interdependent privacy and privacy threat modeling.

\subsection{Interdependent privacy}

IDP captures the networked characteristics of privacy-related decisions~\cite{marwick2014networked}. Owing to this networked nature, the privacy of individuals is bound to be affected by the actions of others, e.g., Facebook users sharing the data of their friends~\cite {DBLP:conf/fc/BiczokC13}. 
IDP emerges from different types of data-sharing scenarios. Profile attributes of a social network user may be harvested via their friends~\cite{DBLP:journals/compsec/SymeonidisBSPSP18}. The location privacy of certain individuals may be threatened by sharing co-location information~\cite{DBLP:journals/tmc/OlteanuHSHH17}. Photo sharing may affect the privacy of friends and bystanders captured in the photo~\cite{DBLP:conf/ndss/OlteanuHDH18}. Even the genetic profile of an individual and associated inferrable medical information might get exposed by an eager relative (i.e., kin genomic privacy)~\cite{DBLP:conf/ccs/HumbertAHT13}. 

A common trait among the aforementioned scenarios is that the affected individual (suffering the privacy loss) is usually unaware, does not give their consent, and is never in control of their own privacy. Another common characteristic is that there is usually no attacker or malicious intent on behalf of the fellow user or friend. (Note that the initiator of the data collection could have malicious or at least selfish intent; see, e.g., the Cambridge Analytica case~\cite{DBLP:journals/compsec/SymeonidisBSPSP18}). Furthermore, data protection regulations do not explicitly recognize IDP; although related notions such as ``amateur controllers''~\cite{helberger2010little,DBLP:journals/compsec/SymeonidisBSPSP18} have been discussed, it remains a legal gray zone. Nevertheless, the resulting privacy threats are real and should be accounted for during any sensible privacy threat modeling activity.

\noindent\textbf{Third-party apps and IDP. }
Investigating IDP remains a niche, albeit one with a growing body of work: see Humbert et al.~\cite{10.1145/3360498} for a survey. Recent studies have turned towards emerging technologies such as smart homes~\cite{alshehri2022exploring}, augmented reality~\cite{o2023privacy}, and large language models~\cite{zhan2024beyond}. Nevertheless, one of the most fertile domains for IDP research has been third-party apps, a polypathological case for privacy. The term itself was coined concerning friend permissions in the Facebook app platform~\cite{DBLP:conf/fc/BiczokC13}. The most publicized incident with IDP as the root cause was the Cambridge Analytica debacle~\cite{DBLP:journals/compsec/SymeonidisBSPSP18}. Even other-regarding preferences were quantified in the context of social apps~\cite{pu2016towards} in the form of monetary value that app users place on their friends’ and their own personal information. Their popularity, the pervasiveness of IDP issues across app platforms~\cite{10.1007/978-3-030-93944-1_5}, and the recent emergence of a platform-agnostic technical mitigation approach~\cite{liu2025idpfilter} make third-party apps an ideal showcase for IDP-related threat modeling.

\noindent\textbf{Real-world IDP incidents. }
We present an overview of four well-publicized incidents rooted in IDP issues in Table~\ref{tab:privacy_issues}. These were all related to social apps (Facebook\footnote{Cambridge Analytica: \url{https://shorturl.at/fV7k7}}\footnote{FB Photo Tagging: \url{https://shorturl.at/4kgqq}}, Strava\footnote{Strava: \url{https://shorturl.at/nBOOb}}, and Twitter\footnote{Twitter: \url{https://shorturl.at/6DsAK}}), had significant societal and legal consequences, and resulted in a variety of response measures from service providers.

\begin{table*}[htbp]
\centering
\caption{Overview of Interdependent Privacy Issues Cases}
\label{tab:privacy_issues}
\begin{tabular}{@{}p{2cm}p{4.5cm}p{2.5cm}p{2.6cm}p{3.4cm}@{}}
\toprule
\textbf{Case} & \textbf{Overview} & \textbf{Consequences} & \textbf{Legal Consequences} & \textbf{Response Measures} \\
\midrule
1. Cambridge Analytica & Cambridge Analytica accessed millions of Facebook users' data without their consent, using this data to target political advertising. This case is significant for interdependent privacy because users' data was shared and used in ways they were unaware of and did not consent to, affecting their privacy and the privacy of their network. & Significant public outcry and loss of trust in Facebook. & Facebook fined \$5 billion by the FTC, various legal proceedings. & Facebook made significant changes, including tightening data access protocols, introducing more robust consent mechanisms for data sharing, and conducting audits of existing apps. \\
\midrule
2. Facebook Photo Tagging Function & Facebook's photo tagging feature uses facial recognition to suggest tags, where user A can tag user B in photos, potentially sharing B's presence and activities without B's consent. This highlights interdependent privacy issues as it involves making decisions about someone else's privacy without their knowledge. & Privacy concerns and debates about consent and data protection. & Lawsuits filed and Facebook eventually settled for \$650 million. & Facebook allowed users to disable facial recognition technology and required clearer consent for using such features, enhancing user control over their personal data. \\
\midrule
3. Strava Heatmap & Strava released a heatmap visualizing all the activities of its users, inadvertently exposing sensitive locations like military bases. This case is an example of interdependent privacy because it involved the disclosure of information about groups of individuals (e.g., military personnel) without their explicit consent. & Potential threats to national security and the privacy of military personnel. & No significant legal consequences, but scrutiny from defense and security agencies. & Strava made changes to its privacy settings to allow users more control over what data is shared publicly and initiated more stringent reviews of what data gets visualized on heatmaps. \\
\midrule
4. Twitter Platform Misuse  & On Twitter, users were able to upload and monetize sexually explicit photos and videos of other individuals without their consent. This represents an interdependent privacy issue as it involves non-consensual sharing of personal content. & Violation of personal privacy, widespread criticism, and distress among victims. & Lawsuits and calls for stricter enforcement of digital content sharing laws. & Twitter implemented more robust mechanisms to detect and prevent the sharing of non-consensual content, including better reporting tools for users to flag inappropriate content and stricter enforcement of community standards. \\
\bottomrule
\end{tabular}
\end{table*}

\subsection{Privacy threat modeling}
Privacy analysis models are critical tools for identifying and mitigating privacy risks in system design and operation. They provide structured approaches to assess potential threats to personal data and ensure compliance with privacy regulations. This section briefly reviews general privacy frameworks and threat modeling methods.

\noindent\textbf{NIST Privacy Framework. }The NIST Privacy Framework provides a structured approach to managing privacy risks. It is designed to be compatible with the NIST Cybersecurity Framework, enhancing its applicability to both security and privacy domains. The framework consists of five functions: Identify, Govern, Control, Communicate, and Protect. The NIST Privacy Framework emphasizes the integration of privacy risk management into organizational processes, promoting a holistic approach to privacy protection that aligns with broader risk management strategies\cite{hiller2017privacy}.

\noindent\textbf{ISO/IEC 29100. }
The ISO/IEC 29100 standard provides a high-level framework for privacy protection. It defines a set of privacy principles and controls to guide organizations in protecting personal data. Key principles include: Consent and choice. Collection limitation. Data minimization. Use, retention, and disclosure limitation. Accuracy and quality. Openness, transparency, and notice. Individual participation and access. Accountability.
ISO/IEC 29100 offers a comprehensive set of guidelines that can be adapted to various organizational contexts, promoting a consistent approach to privacy risk management\footnote{\url{https://www.iso.org/obp/ui/en/#iso:std:iso-iec:29100:ed-2:v1:en}}.

\noindent\textbf{LINDDUN. }
LINDDUN provides a structured methodology for identifying privacy threats through the creation of Data Flow Diagrams (DFDs), which map out data processing activities and data flows within a system. This visual representation helps in systematically identifying potential privacy threats at various points in the data processing lifecycle. The model includes guidance on mitigating identified threats, making it a comprehensive tool for privacy threat analysis.
One of the key strengths of LINDDUN is its comprehensive nature, it considers a broad range of privacy concerns beyond unauthorized access, including threats related to how data is linked and tracked across different contexts and systems. By focusing on these dimensions, LINDDUN helps organizations ensure compliance with privacy regulations and standards while fostering a privacy-aware design culture.
While the LINDDUN framework offers a comprehensive approach to privacy assessment, its shortcomings are also obvious. One notable drawback is its complexity; the detailed analysis required can be time-consuming and may demand significant expertise in both privacy and the specific system architecture being analyzed. This complexity can lead to difficulties in practical application, especially for organizations lacking specialized knowledge in privacy engineering.
Additionally, LINDDUN's traditional focus is heavily theoretical, which can make it challenging to apply in dynamic, real-world environments where quick decisions and agile responses are often necessary. The framework might also overlook some newer privacy issues or emerging threats that arise with technological advancements, as it is not always updated promptly to reflect the latest trends and tools in privacy and data protection.
LINDDUN Go was developed as a response to some of these challenges~\cite{DBLP:conf/eurosp/WuytsSJ20}. It is a more streamlined version of the original LINDDUN framework, designed to be more accessible and applicable to a broader range of users, including those without deep technical knowledge of privacy or security. LINDDUN Go simplifies the process of conducting privacy threat modeling by providing ready-to-use templates, checklists, and guidelines that help identify and address privacy threats more efficiently.
Additionally, Sion et al. proposed an extension to LINDDUN that shifts the focus from a static analysis of data flows to a more dynamic analysis of user-system interactions\cite{DBLP:conf/eurosp/SionWYLJ18}. This approach is particularly effective in identifying privacy threats that emerge from real-time interactions between users and systems, which the original LINDDUN may overlook. In another related effort, Sion et al. proposed a data subject-aware privacy risk assessment model~\cite{sion2019privacy}, which brings PTM closer to GDPR's notion of privacy risk, focusing on the fundamental rights of data subjects. Note that this model enriches the traditional DFD with additional risk inputs.

\noindent\textbf{Emerging approaches. }
While the evolving LINDDUN is easily the most established PTM mechanism, some novel ideas have also emerged in recent years. Adopting a different point of view, xCOMPASS is a more lightweight PTM framework created to improve the scaling and applicability properties of PTM~\cite{maps_xcompass}. Based on attacker personas, xCOMPASS can leverage existing privacy risk frameworks (such as LINDDUN or the NIST Privacy Assessment Framework~\cite{hiller2017privacy}) to provide an operational guide regarding threat actors, threats, and resulting harms.
MITRE's PANOPTIC~\cite{katcherpanoptic} takes a different approach by emphasizing context, i.e., relevant aspects of the socio-technical environment, in addition to privacy-related ``activities'', i.e., potential attacks. A privacy threat in PANOPTIC is a combination of particular threat actions and their corresponding contextual elements. The model encompasses both
actions and inactions, accounts for both benign and malicious intent and acknowledges the system under study as a potential threat agent in addition to external adversaries.

\section{IDP threats: challenges and shortcomings of current methods}
\label{sec: shortcomings}

\subsection{IDP threats: challenges}

\noindent\textbf{User behavior. }
The traditional view on privacy does not hold users accountable for privacy breaches; on the contrary,  IDP scenarios challenge this notion as users' actions can significantly impact the privacy of others. Many users are not fully aware of the implications of their data-sharing actions, especially in terms of the potential exposure of others' data. It is crucial but challenging to educate users about the consequences of their actions, particularly in how they affect others' privacy through data sharing.

\noindent\textbf{Balancing functionality with privacy. }
For system owners, like Strava, data sharing among users forms a fundamental part of the service provided. It is impractical to completely disable these features as they are central to the user experience and the service's value proposition. Finding a balance where user engagement and interaction are maintained without compromising privacy requires careful design and feature implementation.

\noindent\textbf{Conflicting privacy duties. }
System owners face a conflict between protecting users’ data privacy and addressing IDP. It is challenging to protect against IDP without processing data in ways that might not be authorized explicitly by all involved parties. More than that, integrating privacy into the system from the ground up is essential but difficult, especially when user interactions dynamically generate data relationships.

\noindent\textbf{Lack of judicial and enforcement authority. } 
System owners do not possess legal or enforcement powers, which limits their ability to implement stringent policies or interfere in user behaviors directly, especially in gray areas. Crafting policies that effectively address IDP without overstepping user freedoms or system capabilities is a delicate balance. Besides, without enforcement powers, systems must rely on user compliance with policies, which may not always be sufficient.

\subsection{IDP threats: shortcomings of current methods}

As mentioned before, besides its many strengths, LINDDUN, the \emph{de facto} standard for privacy threat modeling, has some limitations concerning its complexity and usability in dynamic real-world systems.

Here we focus on IDP-specific limitations that exist on top of its general shortcomings. Note that although we concentrate on LINDDUN, the IDP-specific shortcomings of other PTMs are also similar. In fact, the organizing principles of existing PTMs (i.e., the threat categories of LINDDUN, the attacker personas of xCOMPASS, and the contextualized threat action of PANOPTIC) are orthogonal to the threats emerging from interdependent privacy.

\noindent\textbf{Individual privacy vs. interdependent privacy. }
LINDDUN focuses on threats to users directly interacting with a system, but not on how their actions can impact third parties. For instance, when someone uploads a contact list to a social media platform, they may consent, but the people on the list have not. LINDDUN may overlook these kinds of indirect privacy risks. In a related effort, 
The concept of interdependent privacy threats is not merely theoretical but has practical implications in real-world systems. As noted by Colesky et al.~\cite{colesky2016critical} in their work on privacy engineering, further explored in the comprehensive analysis by Solove and Schwartz~\cite{solove2020information} on information privacy law, and recently investigated from the control perspective by Alashwali~\cite{alashwali2025}, there is a growing recognition that privacy models need to evolve beyond traditional frameworks to adequately address these interconnected privacy dynamics.  These studies argue for a more holistic approach to privacy threat modeling that includes the potential harms to third parties as a fundamental element of privacy risk assessment.

\noindent\textbf{Forced categorization. }
The LINDDUN model forcibly defines seven major categories (such as Linkability or Identifiability) from the beginning, and then tries to mechanically fit the privacy threats encountered in practice into these categories. A user uploads their contact list (names, phone numbers, emails) to a social media platform or app to find friends or receive recommendations. While the user consents to this, third parties (the contacts themselves) often have no knowledge or control over the sharing of their personal data. Although in the threat tree given by LINDDUN contact lists are placed in linkability\footnote{\url{https://downloads.linddun.org/linddun-trees/tree-full/v241203/Linking.pdf}}, analysts can also classify it into other categories. First, identifiability: unregistered contacts might have their information (e.g., phone or email) matched against existing platform data, enabling the inference of their real identities. Second, disclosure of information: the uploaded data is stored by the platform and may be used in friend suggestions or be visible to others. Third, unawareness: the contacts being uploaded are neither notified nor have they consented to this data usage. When using LINDDUN, analysts are compelled to artificially split a holistic problem into multiple isolated fragments to conform to the model's structure. Specifically, when analysts deal with highly cross-categorical IDP cases, these blurry definitions require analysts to perform significant interpretive work just to fit real-world threats into pre-defined categories, undermining the model’s objectivity and usability.

\noindent\textbf{Non-repudiation is not a threat. }
In the context of IDP, the role of non-repudiation changes fundamentally. In IDP scenarios, privacy is violated due to the actions of another party. Here, non-repudiation does not represent a mechanism of harm, but rather a means of assigning responsibility after the harm has occurred. For example, if user A uploads contacts that include user B’s information, non-repudiation mechanisms ensure that no one can prove A performed the action. In the context of IDP threats, non-repudiation is not a privacy threat in itself, but rather a mechanism for addressing the consequences of privacy violations. It serves as a post-incident accountability tool rather than a proactive defense for privacy protection.

\noindent\textbf{Scalability and limited usability. }
Another fundamental limitation of the LINDDUN model in the context of IDP lies in its poor scalability and limited usability for analyzing complex, multi-actor data flows. In IDP scenarios, the privacy risk is clearly centered around one user's actions affecting another's privacy, often without their awareness or consent. These risks are not isolated; they are inherited and propagated across the DFD. A single action, such as uploading contact data, can lead to multiple downstream consequences: the data may be forwarded to other components, analyzed by a profiling algorithm, or stored in a database. Each step introduces new, derivative privacy threats.
However, LINDDUN requires threat modeling to be performed on each element of the DFD independently, without a built-in mechanism for tracing inherited or propagated risks. This results in a high degree of manual, repetitive work. Analysts must redundantly assess the same privacy concern (e.g., unauthorized data exposure) across multiple nodes, even when the source of the risk originates from a single upstream action. For example, if contact data uploaded by one user is processed by several backend services, each step must be evaluated in isolation under LINDDUN, even though the privacy impact remains fundamentally tied to the original upload.
This fragmented approach not only creates unnecessary workload but also makes it difficult for analysts to maintain a coherent picture of interdependent risks. LINDDUN lacks the abstraction tools necessary to model cascading privacy effects, which are essential in understanding how a single user's action can affect many others through a system.

\section{Guiding principles for IDP-specific threat modeling}
\label{sec:principles}

\subsection{Elicitation and analysis: the 3R principles}

Kamleitner and Mitchell proposed the 3R framework for addressing interdependent privacy infringements\cite{kamleitner2019your}.
\begin{enumerate}
\item \textbf{Realize data transfer:} This principle emphasizes the importance of awareness regarding the flow of data within and between systems. It inspires organizations to develop better tracking and management systems that make data transfers transparent and traceable, helping stakeholders understand how their information might be shared or exposed.\item \textbf{Recognize others’ rights:} By focusing on recognizing the rights of all individuals affected by data transactions, this principle encourages a broader consideration of privacy impacts. It promotes the adoption of privacy measures that consider the rights of secondary data subjects, fostering a more inclusive approach to privacy.
\item \textbf{Respect others’ rights:} This principle underscores the ethical imperative to not only acknowledge but actively respect the rights of others by implementing privacy protections that prevent harm. It calls for privacy practices that are not just compliant with legal standards but are also fundamentally aligned with respecting individual privacy rights, thus ensuring that data practices do not infringe on the privacy of others.
\end{enumerate}

Although the 3R principles provide inspirational value in understanding and handling interdependent privacy concerns, their practical limitations in complex digital ecosystems are also clear.

\begin{enumerate}
    
    \item \textbf{Challenges in operationalization:} The principles require organizations to operationalize complex processes of tracking data flows and managing consents that may involve multiple stakeholders with varying expectations and legal rights. Implementing such comprehensive systems can be technically and administratively challenging.

    \item\textbf{ Dynamic and scalable implementation:} As digital environments evolve, the static nature of some privacy frameworks may struggle to keep pace. The principles should be adaptable to handle dynamic interactions and scalable to accommodate growing data ecosystems, which might require continuous updates to privacy practices and technologies.

    \item \textbf{Balancing rights with business needs:} There might be tensions between respecting privacy rights and pursuing business objectives, such as data analytics and monetization strategies. Organizations need to find a balance that respects privacy without stifling innovation, which can be a delicate and complex task.
\end{enumerate}

\begin{table*}[htbp] 
\centering
\caption{Alignment of IDP Event Response Measures with 6A Principles}
\label{tab:6A_alignment}
\begin{tabularx}{\textwidth}{>{\raggedright\arraybackslash}X *{6}{>{\raggedright\arraybackslash}X}}
\toprule
\textbf{IDP Event} & \textbf{Awareness} & \textbf{Authorization} & \textbf{Access} & \textbf{Accountability} & \textbf{Auditability} & \textbf{Alignment} \\
\midrule
Cambridge Analytica & Increased transparency; policies revised for clarity. & Enhanced consent mechanisms. & Restricted data access. & Legal proceedings ensure responsibility. & Audits of app developers. & Policies revised to align with user expectations. \\
\addlinespace
\midrule
Facebook Photo Tagging Function & Users informed about facial recognition. & Opt-out available for tagging. & Controlled tag suggestions. &  &  & Adjustments align with privacy concerns. \\
\addlinespace
\midrule
Strava Heatmap & Publicized privacy options. & Users control data sharing. & Sensitive locations hidden. & Scrutiny from security agencies. & Reviews of shared data visualizations. & Features redesigned to align with security needs. \\
\addlinespace
\midrule
Twitter Platform Misuse &  &  &  & Enforcement of stricter content rules. & Strengthened manual review and improved response speed. &  \\
\bottomrule
\end{tabularx}
\end{table*}

\subsection{Analysis and mitigation: the 6A principles}

Based on the analysis of IDP incidents, we have summarized the new 6A principles that are more operationally instructive in practical applications. The 6A principles—Awareness, Authorization, Access, Accountability, Auditability, and Alignment provide a structured approach to safeguarding interdependent privacy in digital environments. These principles focus on enhancing user understanding, ensuring proper permissions, and controlling data access.

\begin{enumerate}

\item \textbf{Awareness. }This principle emphasizes the importance of making users fully aware of the data collection processes and how their data may be used, especially in scenarios where it might affect other individuals. Raising awareness involves clear communication about the nature of interdependencies in data privacy, such as how data shared in one context might be linked or combined with data from other sources, affecting multiple parties. Providing users with comprehensive information enhances their understanding and ability to make informed decisions about their data.

\item \textbf{Authorization.} Authorization ensures that consent is obtained from all individuals whose data is collected or processed. This is crucial in managing interdependent privacy as it involves not just the primary user but also any secondary individuals whose information might be indirectly involved. The principle of Authorization requires that consent mechanisms are clear and explicit, and cover the breadth of data uses, particularly in how they may impact the privacy of others beyond the primary data subject.

\item \textbf{Access. }Controlling access to data is essential to prevent unauthorized or unintended use that could compromise privacy. This involves implementing stringent access controls and authentication mechanisms to ensure that only authorized individuals or entities can view or process data. In the context of interdependent privacy, controlling access is vital to protecting the interconnected data of users, thereby limiting exposure and reducing the risk of privacy breaches affecting multiple individuals.

\item \textbf{Accountability. }While the current principles cover the proactive aspects of privacy management, Accountability would address the reactive measures. This involves setting up mechanisms to monitor compliance with the privacy policies and the actions taken when breaches occur. It ensures that there are clear responsibilities and repercussions in place to address any misuse of data or violation of policies.

\item \textbf{Auditability. }Auditability could ensure that there are mechanisms to verify and review how data is accessed and used over time. This would help in maintaining transparency and trust, especially in systems where data might affect numerous users. Audit trails that log access and changes to data can help in detecting, investigating, and responding to privacy incidents that could impact multiple individuals.

\item \textbf{Alignment. }This principle could focus on ensuring that all privacy practices align with both legal requirements and user expectations. Alignment involves continuously updating privacy practices to reflect changes in laws, technology, and user perceptions. It also emphasizes the importance of aligning the operational practices with the stated privacy policies to avoid discrepancies that might lead to privacy breaches.

\end{enumerate}

By matching the response measures after four real-world IDP incidents, the guiding significance of the 6A principles can be justified, see Table~\ref{tab:6A_alignment}.

\section{Interdependent Privacy Threat Analysis}
\label{sec:ida}
In this section, we focus on developing the Interdependent Privacy Threat Analysis (IDPA) concept to address IDP threats. IDPA is proposed as a guiding framework for managing novel privacy challenges where user data is not isolated but inherently connected. This means that privacy threat analysis goes beyond targeting one role in the system; it also covers the effects caused by the role. At the same time, IDPA aims to improve on the usability aspect by providing more granular analysis steps, making it easier for practitioners to get started when facing complex IDP threats. To better illustrate each step of the work, we take the ``Find friends via contacts list''  data flow diagram as an example, which is a common function of online social networks. For easy reading, we placed the abbreviation list \ref{tab:abbreviations} at the end of the paper.

When addressing IDP threats, traditional techniques and legal measures designed for individual privacy issues fall short. These threats involve interconnected data among individuals, meaning that the protection of privacy isn't just about the individual but extends to a network of people. Therefore, a bird's-eye view is required.

\subsection{Elicitation: classifying data flows}
\begin{figure}[tb]
        \centering
        \includegraphics[width=0.4\textwidth]{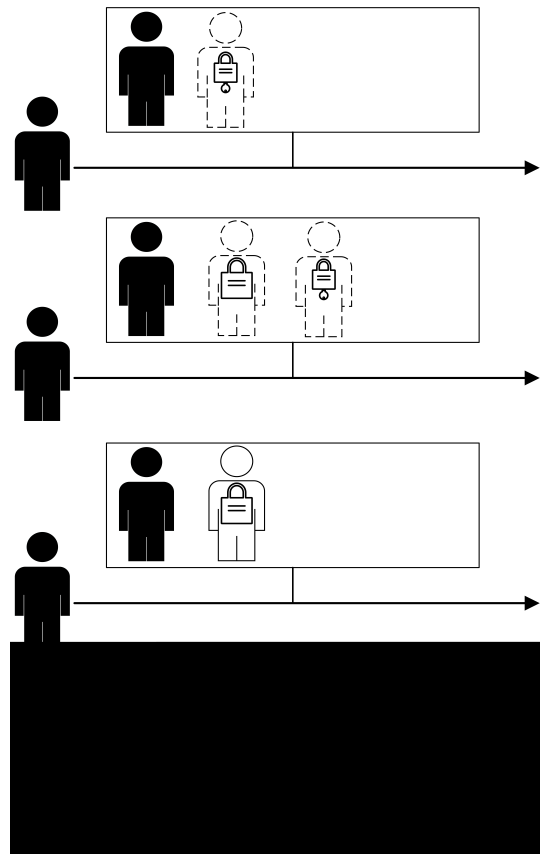}
        \caption{3 types of Data Flow: NIDPF, PIDPF and IDPF}
        \label{fig:IDPdataflow41}
\end{figure}

As the starting point of IDPA, the first step in analyzing IDP threats is to label the data flows. Data flows within a system are categorized into three distinct types, each reflecting the level of interdependency and potential privacy impact. These are
invoking interdependent privacy (IDPF), potentially invoking interdependent privacy (PIDPF), and not invoking interdependent privacy (NIDPF); see Fig.~\ref{fig:IDPdataflow41}. Note that ``data sender'' corresponds to the user initiating the sharing, while ``data stakeholder'' refers to a person whose data is included in the shared data object (could also be a non-user!).

\noindent \textbf{Privacy-Interdependent Data Flow (IDPF). }
This category includes data flows where one entity transmits data that contains information about other users. For instance, when a user uploads their contact list, the data flow contains the personal details of multiple individuals, not just the uploader. Similarly, if an application sends information to a user that includes data derived from other users (e.g., shared ride details in a ride-sharing app), this also falls under IDP. The likelihood factor of the risk of impacting others' privacy is $1$ (once the corresponding data flow is instantiated).

\noindent \textbf{Potentially Privacy-Interdependent Data Flow (PIDPF). }
Data flows are categorized as potentially privacy-interdependent when the data transmitted could possibly include information about other users. For instance, this category is typical in social media interactions, such as when a user posts a photo that may (unintentionally) include other people, or when they send messages that refer to or affect other individuals. The risk of impacting others' privacy exists but its likelihood depends on the actual data objects transmitted.

\noindent \textbf{Normal Data Flow (NIDPF). }
All other data flows that do not necessarily involve or potentially involve information about other users are classified as ``normal'' or NIDP. These are standard data transactions that concern only the data of the user involved in the interaction and do not inherently affect others.

\begin{figure}[tb]
        \centering
        \includegraphics[width=0.4\textwidth]{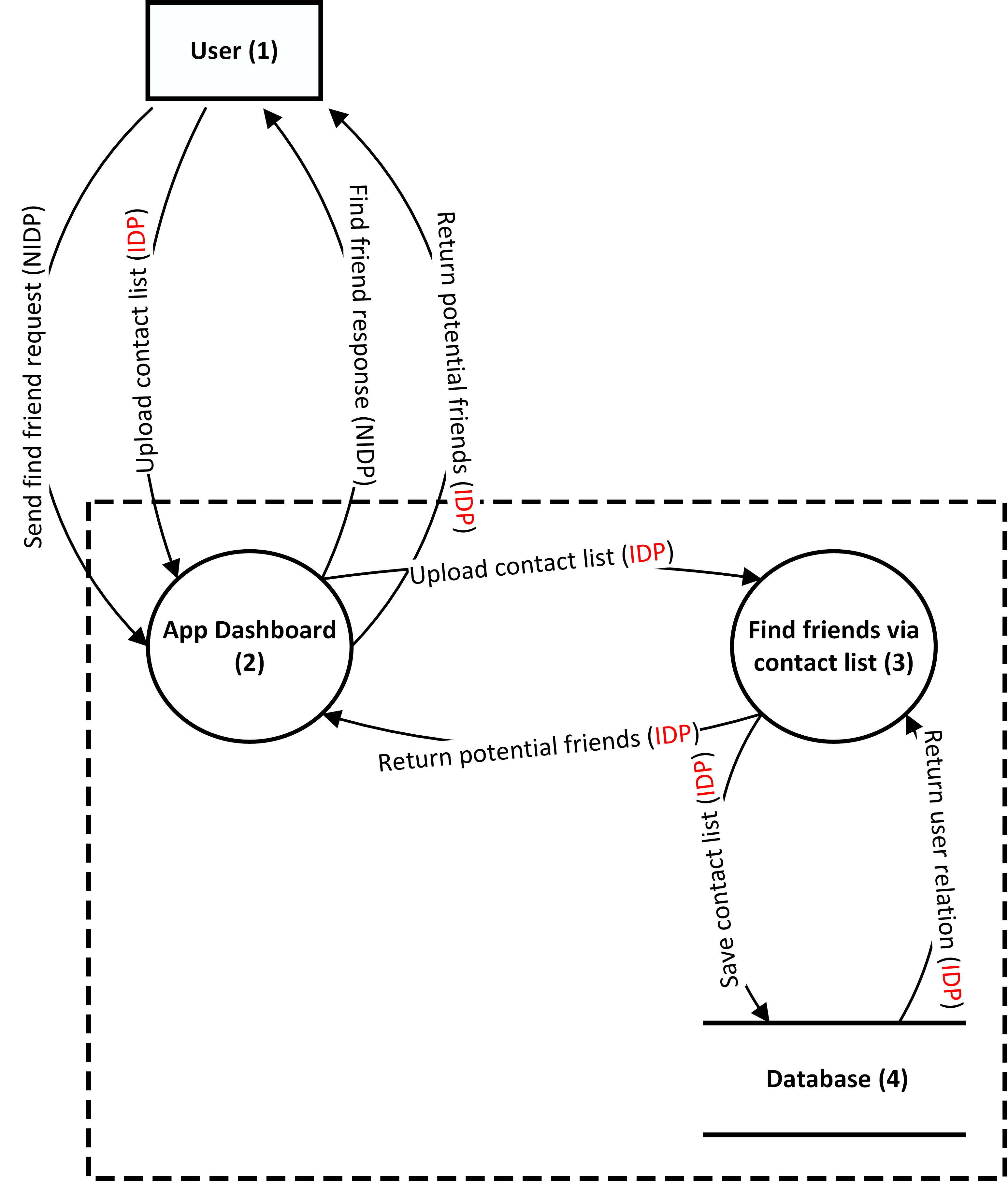}
        \caption{Labeled ``Find friends via contact list'' DFD}
    \label{fig:findfriend}
\end{figure}

We use the previous example to specifically illustrate how to use these three predefined data flow categories. In the ``Find friends via contact list'' DFD, when users upload their contact lists, they share others' information with the WeChat application; therefore, the data flow is marked as IDP (see Fig.~\ref{fig:findfriend}). Similarly, when WeChat stores the contact lists in its database, the data flow is also IDP. 
In a related ``Post pictures'' DFD, when users post their photos, it could potentially contain others' likenesses or other information; therefore, the corresponding data flows are marked as PIDP.

\subsection{Elicitation: threat categories}

The crux of the IDP problem lies in whether an individual or organization holds or transmits the private data of other individuals (without them being aware and/or consenting). When analyzing IDP privacy threats, the original categories of existing models are no longer applicable. Therefore, the second step of IDPA is to introduce three new IDP-specific threat categories.

\noindent \textbf{Improper sharing of privacy-interdependent data. }
Such an action occurs when data that is linked to or related across various entities and individuals is shared without proper authorization, oversight, or compliance with privacy regulations. For example,  a user shares their contact information with the system without the contacts being consenting or even aware. Such an IDP threat usually materializes when data is transferred.

\noindent \textbf{Improper storage of privacy-interdependent data.}
This category refers to the privacy-invasive storage practices that do not adequately protect data linked or related to multiple persons or a single non-user. For example, a user, John, uploads a family group photo to his social media profile and tags his relatives, including some minors and family members who prefer to keep their social presence minimal. While John has consented to the platform's data storage policies by agreeing to their terms of service, the tagged family members may not have directly given their consent for their images or relational information to be stored or processed by the platform.  

Note that mechanisms like Facebook's well-known ``Timeline review'' control (which affects both sharing and storage) fall short of mitigating such threats. A tagged person can only control what appears on their own timeline but has absolutely no means to restrict the post's appearance on other timelines or news feeds. What's more, a tagged non-user doesn't even become aware of the situation.

\begin{table*}[tb]
\centering
\caption{Definitions of awareness, consent, and access control}
\label{tab:ACA}
\begin{tabular}{p{2.2cm} *{3}{p{4.3cm}}}
\toprule

&  Improper Sharing &  Improper Storage & Improper Processing\\
\midrule
Awareness: are all users involved and non-users affected are fully aware of  &
what data is shared, how it is used, and who it is shared with? &
where their data is stored, how long it is kept, and which security measures keep it safe? &
how their data is processed, including any implications for the privacy of other individuals? \\
\midrule

Consent: have all users involved and non-users affected given their explicit consent under clear terms &
for their data to be shared? Do they understand the details of the sharing? &
for their data to be stored? Do they understand the details of the storage, including duration and purpose? &
for their data to be processed? Do they understand the scope, purpose, and potential outcomes of the processing? \\
\midrule

Access Control: does the system ensure that &
only authorized users can share data? This includes setting up permissions that align with the agreed terms of data sharing. &
access to stored data is restricted to authorized stakeholders only? Are proper security measures in place to safeguard the data? &
only stakeholders, who need to process the data for legitimate purposes can do so? \\
\bottomrule
\end{tabular}
\end{table*}

\noindent \textbf{Improper processing of privacy-interdependent data. }
This category captures the issues arising from an entity processing and mining relationships between data points involving multiple users (and potentially non-users), without proper permission. Data analytics might reveal sensitive relationships or information that was not intended to be disclosed, thus infringing on the privacy of individuals involved. Consider a scenario involving a financial service app that utilizes customer transaction data to offer personalized financial advice. The app collects data about financial interactions that often involve other parties, such as family members or business partners.  The app uses an algorithm to analyze spending patterns and suggest budgeting tips. To enhance its services, the app begins to cross-reference user data with that of other users to identify common spending trends and offer group discounts. While doing so, it (inadvertently) processes data in a way that reveals personal spending habits and financial relationships between users and non-users (e.g., family members who share credit cards but aren't app users themselves).

\subsection{Analysis: awareness, consent, and access control}

Mapping threats to categories and conducting an actual PTM process has its documented challenges. Owing to the specific nature of IDP threats, we can streamline analysis by focusing on three foundational dimensions and checking whether their key requirements are satisfied by the system under study: 
 
awareness, consent, and access control (see Table~\ref{tab:ACA}).

\noindent \textit{Awareness. }
Being a first-step enabler of privacy notwithstanding, awareness is often imperfect, as individuals often lack information on how their data or data they are implicated in, is being used or shared across different platforms and entities. In scenarios where data is interconnected, the actions of one user can significantly impact the privacy of another. By fostering a high level of awareness among all stakeholders, individuals can make informed decisions about their personal data, understand the potential repercussions of data sharing, and act to protect their privacy rights more effectively.

\noindent \textit{Consent. }
A foundation of the GDPR, consent is critical in ensuring that all natural persons involved in data exchange have agreed to how their data is being collected, used, and shared. Consent must be informed, explicit, and freely given, aligning with stringent privacy regulations. Establishing robust consent mechanisms ensures that all individuals have a say in how their data is handled, thereby upholding both legal requirements and ethical standards.

\noindent \textit{Access Control. }
Access Control needs to be rigorously implemented to restrict access to data and operations to authorized stakeholders only. This control is vital to prevent unauthorized use or disclosure of sensitive information, particularly in environments where data is shared across platforms and among multiple users. Effective access control mechanisms help in mitigating data breaches and ensuring that only stakeholders with legitimate reasons and permissions can access or use the data. Note that proper authorization policies are a prerequisite for effective access control.

\subsection{Mitigation: ensuring accountability, auditability, and alignment}

The common approaches of raising awareness, seeking consent, and controlling access, although foundational, often fall short when dealing with the complex nature of IDP issues. In scenarios where user actions can inadvertently affect others, such as in the cases of Twitter misuse, these preventive measures do not address the broader implications of data interactions that span across multiple users and systems. Tables~\ref{tab:privacy_issues} and ~\ref{tab:6A_alignment}The analysis in Section~\ref{sec:case} has proven that traditional measures preventing the leakage of private information are not always feasible in the case of IDP. Therefore, whether to formulate responsibility identification rules, accountability measures, and compliance with the rules has become equally important countermeasures. 
In line with the 6A principles, auditability ensures the establishment of a transparent mechanism to review and monitor how data is used and shared; so that any abuse or violation can be traced back to its source. Accountability ensures that such violations are punished, thereby preventing negligent or malicious behavior. The significance of accountability in IDP threats lies in the division of responsibilities and informing users of their responsibilities when the system is unable to interfere with user behavior. Alignment helps the system check whether it has complied with its own regulations on IDP.

\begin{figure*}[!ht]
\centering
\includegraphics[width=\textwidth]{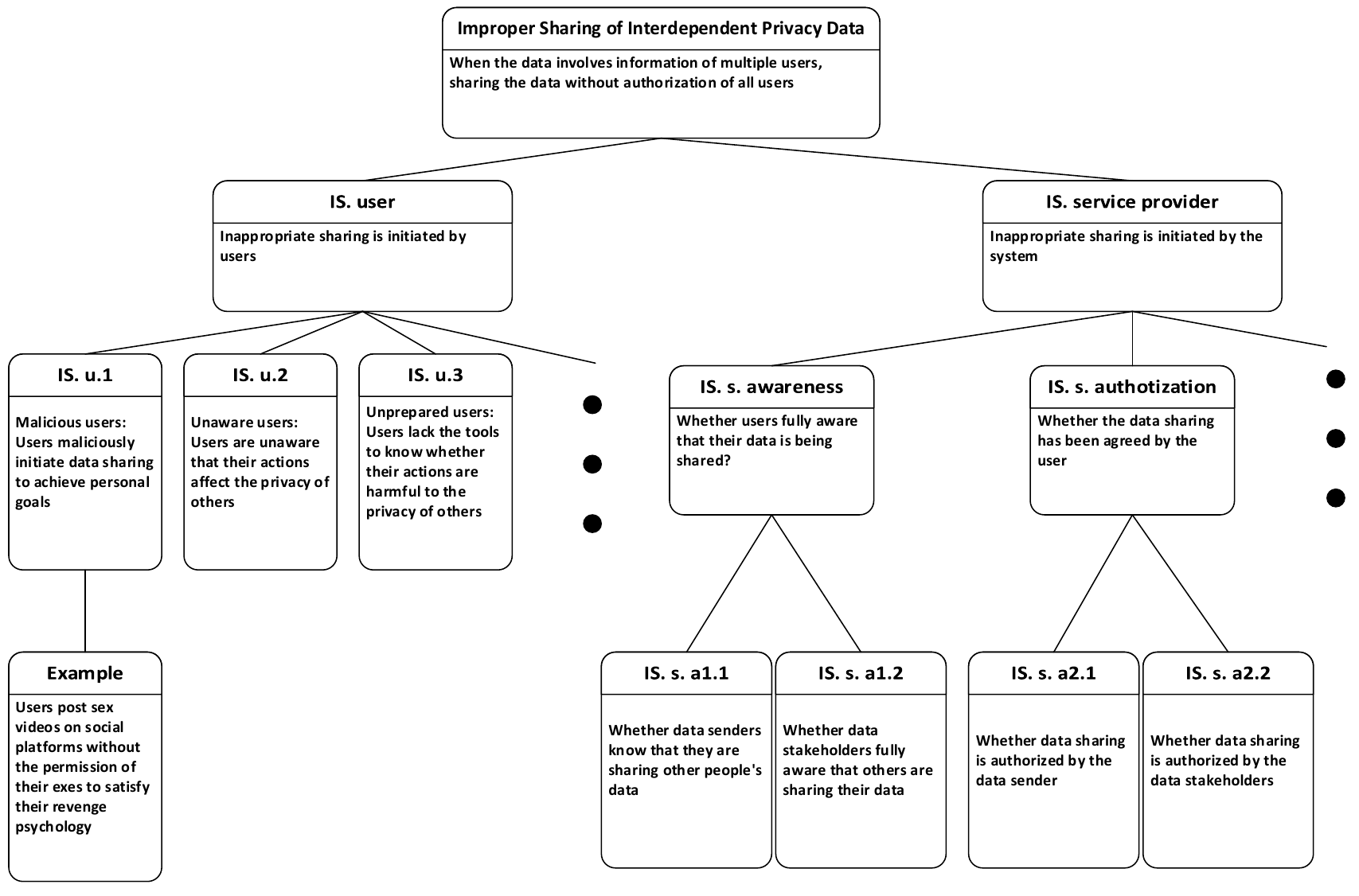}
\caption{IDPA threat tree example}
\label{fig:threat tree}
\end{figure*}

\subsection{Mitigation: misactors and tailored mitigation strategies}

In the context of IDP threats, differentiating potential misactors, both intentional and unintentional, is critical. This analysis should extend to different user types within the system, ranging from naive users who may unknowingly cause data breaches to insider threats where users intentionally exploit system vulnerabilities. In scenarios involving IDP threats, the actions of one user can inadvertently affect the privacy of others. For instance, when a user shares a document containing collaborative input from multiple people, the system has limited control over the sharing decisions of that user. Furthermore, traditionally, it is the system’s responsibility to safeguard user privacy and manage data in a manner that adheres to legal and ethical standards. Users, typically, should not be held responsible for systemic failures that lead to privacy breaches. This principle is crucial because users generally lack the technical knowledge and control over how systems manage and protect stored data. Although users should not be held accountable for systemic privacy threats, 
they still pose IDP threats; analyzing their behavior can aid in crafting effective mitigation strategies. For example, if it is observed that users frequently share sensitive information in certain contexts, the system can introduce targeted educational campaigns or system prompts to encourage safer data handling practices.

\noindent \textbf{Malicious Users (MU). }
These are individuals or groups who intentionally exploit IDP-related weaknesses to access or steal personal information. They may use sophisticated methods to bypass security measures and gain unauthorized access to data.

\noindent \textit{Mitigation. } 
\begin{itemize}

\item Implement advanced security measures: use encryption, anomaly, and intrusion detection systems to protect data from unauthorized access.
\item Security Audits: conduct periodic audits, including penetration testing, to identify and address weaknesses.
\end{itemize}

\noindent \textbf{Indifferent Users (IU). }
These are individuals or entities who may not necessarily have malicious intent but display a disregard for the privacy of others. Their (in)actions may inadvertently lead to privacy breaches or the exposure of sensitive information.

\noindent \textit{Mitigation. }
\begin{itemize}
\item Privacy awareness training: conduct regular training sessions to educate about the importance of privacy and the consequences of negligence.
\item Implement privacy-by-design: re-design systems and processes that inherently protect privacy by limiting data exposure and enhancing user control over personal information. 
\end{itemize}

\noindent \textbf{Unprepared Users (UU). }
These actors want to protect others' privacy but lack the necessary tools or knowledge to do so effectively. Their attempts might be undermined by acting along inadequate or outdated privacy practices.

\noindent \textit{Mitigation. } 
\begin{itemize}
\item Provide access to privacy tools: offer easy-to-use privacy tools that can help in protecting (others') personal data~\cite{liu2025idpfilter}.
\item Guidance: Distribute privacy guidelines and best practices.
\end{itemize}

\noindent \textbf{Uninformed Users (UFU). }
These are individuals who are not aware of how their actions may violate the privacy of others. They may unknowingly share or process data in ways that infringe on privacy due to a lack of awareness about (interdependent) privacy norms and regulations.

\noindent \textit{Mitigation. }
\begin{itemize}
\item Educational campaigns: launch campaigns to increase awareness about privacy rights and responsibilities.
\item Clear privacy policies: ensure that privacy policies are comprehensive but easy to understand.
\end{itemize}

\noindent \textbf{Service Providers (SP). }
These entities process and manage data as part of their business operations. They may unintentionally become vectors for privacy threats if their systems are not adequately secured or if they fail to enforce strong data protection policies.

\noindent \textit{Mitigation. }
\begin{itemize}
\item Strict data handling protocols: enforce rigorous data handling protocols and regular compliance checks.
\item Data protection agreements: require strict data protection terms in agreements with third parties.
\end{itemize}

\noindent\textbf{Government authorities (GA). }
Governmental organizations might access or request data for regulatory or judicial purposes, but can also contribute to privacy threats if such actions are not transparent or if the personal data is not handled according to strict protection standards.

\noindent \textit{Mitigation. }
\begin{itemize}
\item Regulatory compliance: adhere to data protection regulations and ensure that any data requests are fulfilled only if fully compliant with legal standards.
\item Transparency and accountability: maintain high levels of transparency regarding data requests and implement robust accountability mechanisms to oversee data handling.
\end{itemize}

\begin{table}[]
\centering
\resizebox{\columnwidth}{!}{%
\begin{tabular}{@{}llllllcc@{}}
\toprule
\multicolumn{6}{c}{Data Flow} &
  \multicolumn{1}{l}{Misactor} &
  \multicolumn{1}{l}{Privacy Threat} \\ \midrule
Source &
  Flow &
  Destination &
  IDPF &
  PIDPF &
  NIDPF &
  MU &
  IS \\
User (1) &
  \multicolumn{1}{c}{upload contacts} &
  App (2) &
  \multicolumn{1}{c}{X} &
  \multicolumn{1}{c}{} &
  \multicolumn{1}{c}{} &
  X &
  X \\ \bottomrule
\end{tabular}%
}
\caption{Threat mapping example}
\label{tab: mapthreatsexample}
\end{table}
The classification of data flows, identification of potential types of IDP threats, and precise detection of misactors can be illustrated by the examples in Table~\ref{tab: mapthreatsexample} and Fig.~\ref{fig:threat tree}.

\section{Case Study: WeChat}
\label{sec:case}

In this section, we analyze the WeChat super-app using the IDPA framework. Our goal is to validate IDPA in modeling interdependent privacy threats. WeChat\footnote{\url{https://www.wechat.com/en/}} is a multifunctional social media app developed by the Chinese company Tencent. Launched in 2011, it has become one of the world's largest standalone mobile apps, with over a billion monthly active users. Primarily known for its messaging and calling features, WeChat actually offers more, including social networking, mobile payments via WeChat Pay, and a platform for businesses and brands to interact with customers. Users can send text messages, make voice and video calls, share images and videos, and post to a timeline-like feature called ``Moments''. WeChat is an integral part of daily life in China, deeply embedded in both personal communications and commercial activities.

\subsection{System model}
The DFD components of the WeChat application framework are given in Table~\ref{table:DFDelements}. 

\begin{figure*}[t]
\centering
\includegraphics[width=\textwidth]{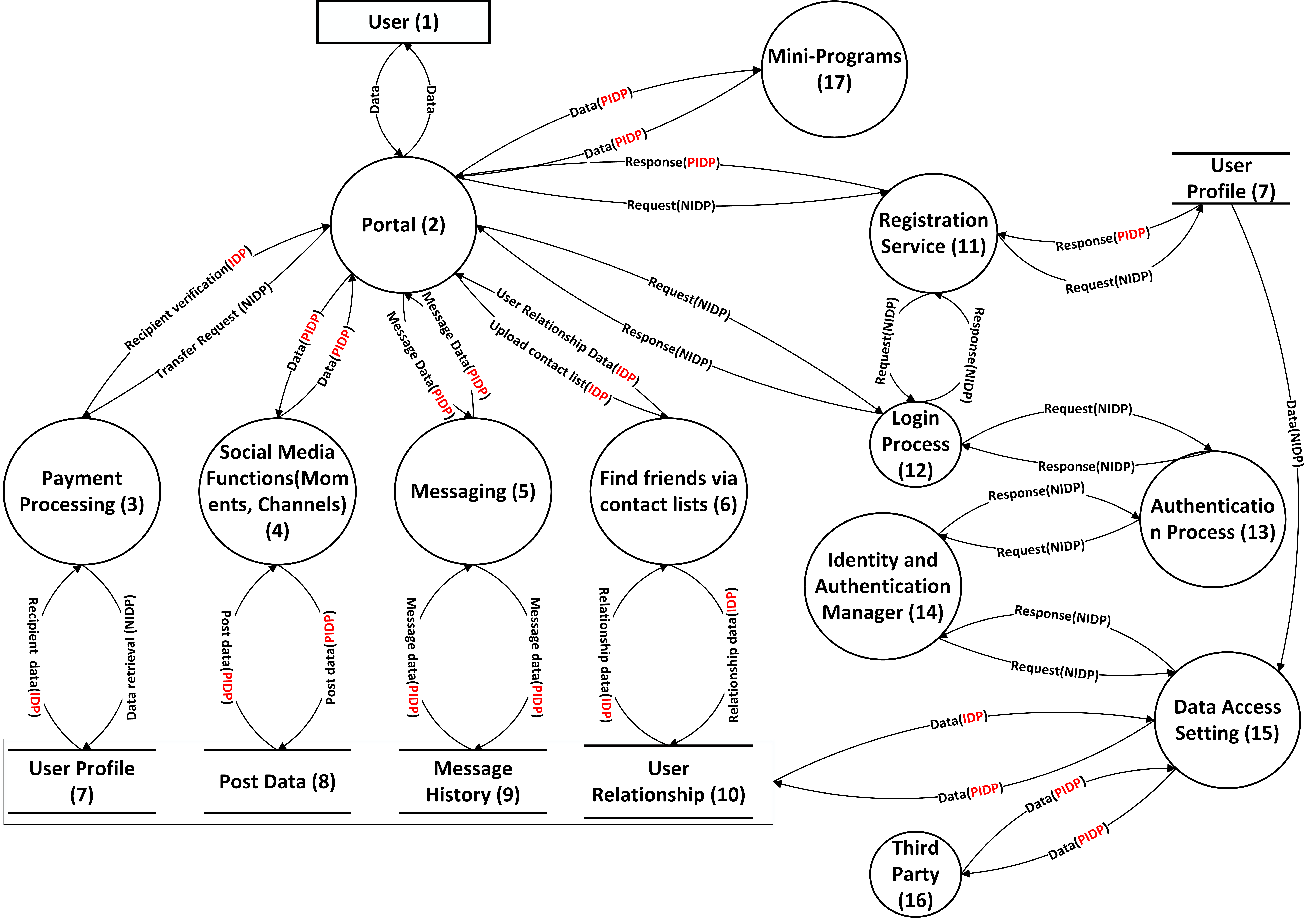}
\caption{WeChat Data Flow Diagram}
\label{fig:wechatDFD}
\end{figure*}

\begin{table}[tb]
\centering
\caption{WeChat application components}
\label{table:DFDelements}
\resizebox{8cm}{!}{\begin{tabular}{p{2cm} p{6cm}}
\toprule
\multicolumn{2}{c}{\textbf{External Entities}} \\
\midrule
User & Interact with the app through various features. \\ 
\midrule
\multicolumn{2}{c}{\textbf{Processes}} \\
\midrule
Messaging & Handles sending and receiving messages. \\
Registration Service & Handles registration service. \\
Login Process & Handles login. \\
Authentication Process & Handles login authentication. \\
Identity and Authentication Manager & Handles login and registration authentication process. \\
Data Access Setting & Handles data access settings made by user. \\
Portal & Handles sending responses and receiving requests. \\
Social Media Functions (Moments, Channels, Live Streaming) & Manages posting, viewing, and interacting with content. \\
Payment Processing & Manages transactions via WeChat Pay. \\
Mini Programs & Facilitates the operation of embedded apps within WeChat. \\
Third-party & Include payment gateways, advertisement services, and external app integration. \\
Find friends via contact lists & Search Friends from Contacts \\
\midrule
\multicolumn{2}{c}{\textbf{Data Stores}} \\
\midrule
User Profile & Stores user personal data, preferences, and activity logs. \\ 
Message History & Archives of all user communications. \\
Post Data & Stores posts, videos, and images shared in social media features. \\
User relationship & Stores social connections\\

\bottomrule
\end{tabular}
}
\end{table}
Note that we did not list all the functions of WeChat due to its complexity. However, we listed the popular functions that matter in the context of (interdependent) privacy threats. Given the components, we draw the (simplified) data flow diagram of the WeChat framework in Fig.~\ref{fig:wechatDFD}.

\subsection{Mapping threats}
We analyzed IDP threats of the WeChat application using IDPA. The analysis is structured into three distinct steps: i) the precise classification of data flows according to their pertinence to IDP and initial threat discovery, ii) mapping to newly defined IDP threat categories, and iii) misactor analysis. The resulting threat map is visualized in Table~\ref{table:map-table}. Note that we do not give a comprehensive description due to the lack of space, but we present key findings below. 
As described in the table, IDPA didn't find many IDP threats during user registration and login; it is easy to see that this process is solely related to the individual user themselves.

\subsection{IDP threats in WeChat}
During our analysis, we observed distinct patterns in how IDP threats manifest within the WeChat application. The analysis revealed three primary areas of concern.


\noindent\textbf{User Registration and Login. }
During the user registration and login phases, interdependent privacy threats are relatively scarce. The main issue arises when a new user attempts to register using specific information that belongs to another person, such as a phone number or email address. In such cases, the system performs a lookup in its database to check if the user is already registered, and if so, it informs the new registrant that the number or email is already in use. This action, although designed to prevent duplicate registrations, inadvertently shares the registration status of a person’s contact information with another user, thereby creating an IDP threat. This threat stems from the system disclosing to the new user that someone else has already registered with those details, potentially without the consent of the original user.

\begin{figure*}
\includegraphics[width=\textwidth]{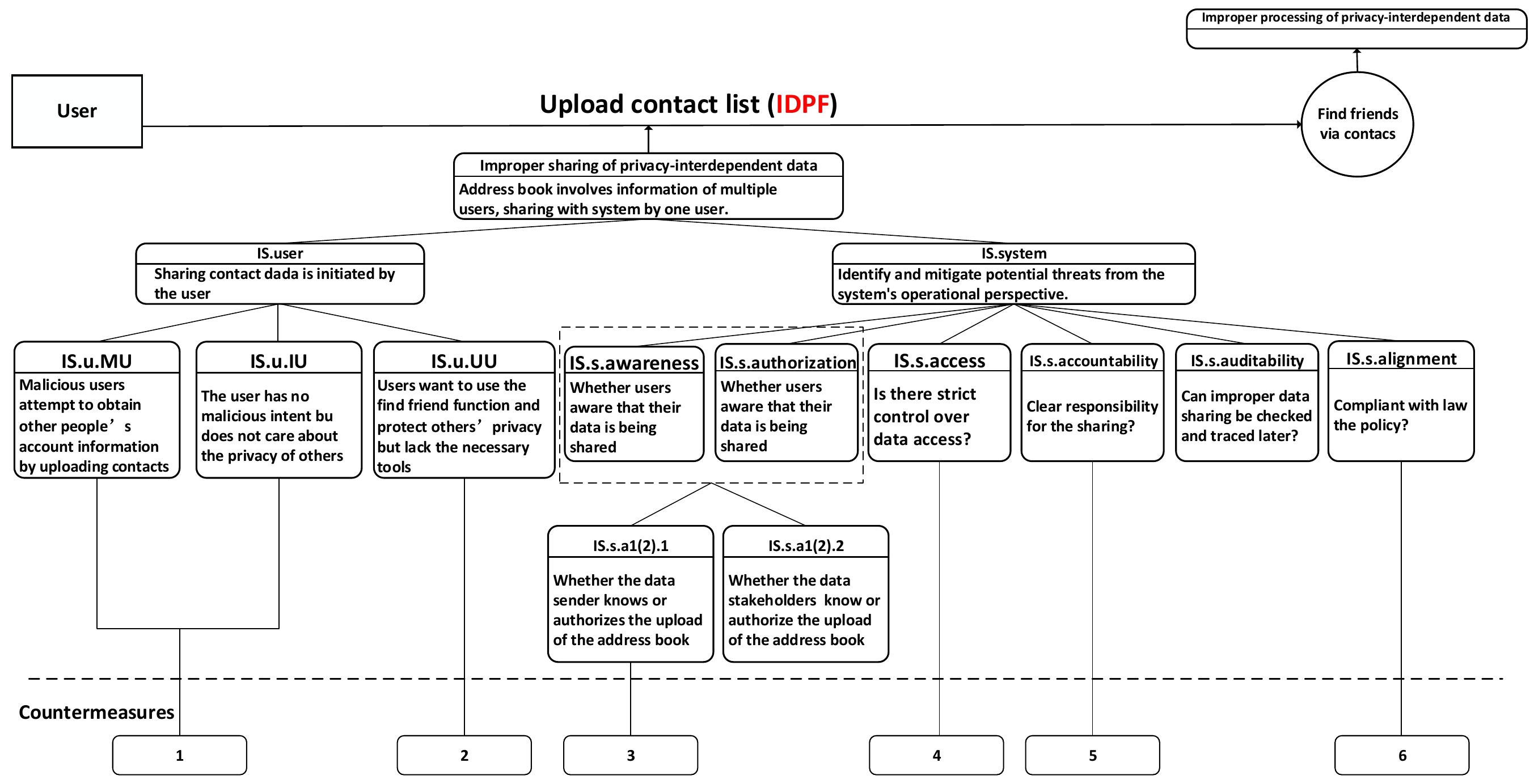}
\caption{Threat tree for ``upload contact list''}
\label{fig: wechatthreattree}
\end{figure*}

\begin{figure}[tb]
        \centering
        \includegraphics[width=0.3\textwidth]{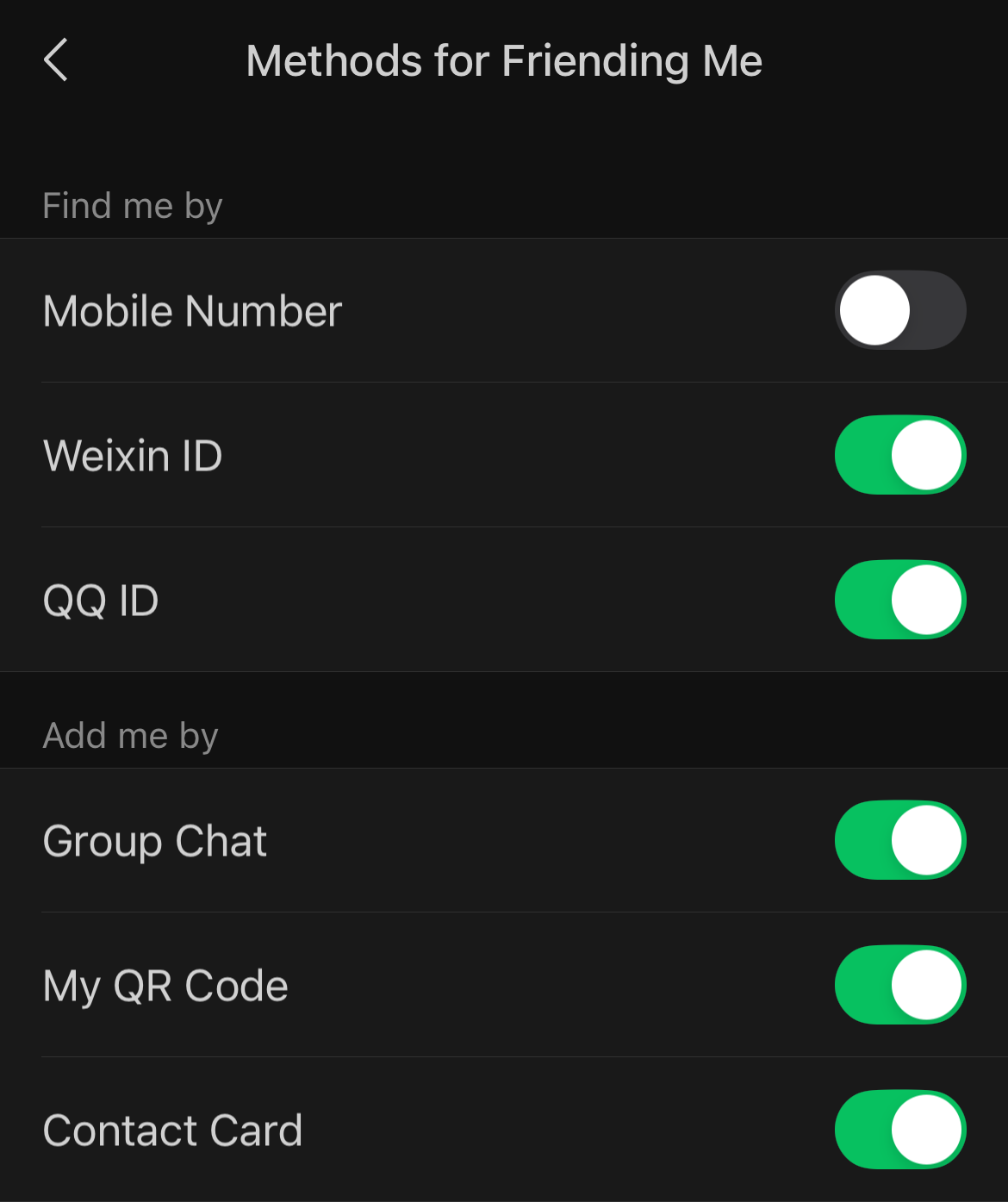}
        \caption{Disable ``find me via phone number''}
        \label{fig:Disable phone}
\end{figure}

\noindent\textbf{Find friend via contacts. }
According to Fig.~\ref{fig:wechatDFD}, data flows $(2) \rightarrow (6)$  show that the user uploads the address book to the application, and the address book is confirmed to contain other people's personal data, so the data flow of uploading the address book is marked as IDPF. When IDPF exists, there must be an IDP threat. Therefore, the second step is to map the threat. This data flow only involves data sharing. According to the threat category definition, it can be easily concluded that this threat is of type IS. We expand this process into a threat analysis tree; the details are shown in Fig.~\ref{fig: wechatthreattree}. This process of sharing other people's data is initiated by the user. From the analysis of user motivation, it can be seen that its misactor types are MU, IU, and UU. ``Find friends via contacts'' is a common function that users need, and WeChat cannot interfere with the users' sharing their address books. In response, WeChat's measures are aimed at MU and IU. Although WeChat cannot prohibit the sharing of address books, it restricts their attempts to obtain other people's account information. The specific measure is 1: all users can prohibit others from finding their accounts through address book information. Although this measure cannot prohibit one's personal information from being uploaded to WeChat, it mitigates the possible consequences of this behavior, as shown in Fig.~\ref{fig:Disable phone}.

\begin{figure}[tb]
        \centering
        \includegraphics[width=0.3\textwidth]{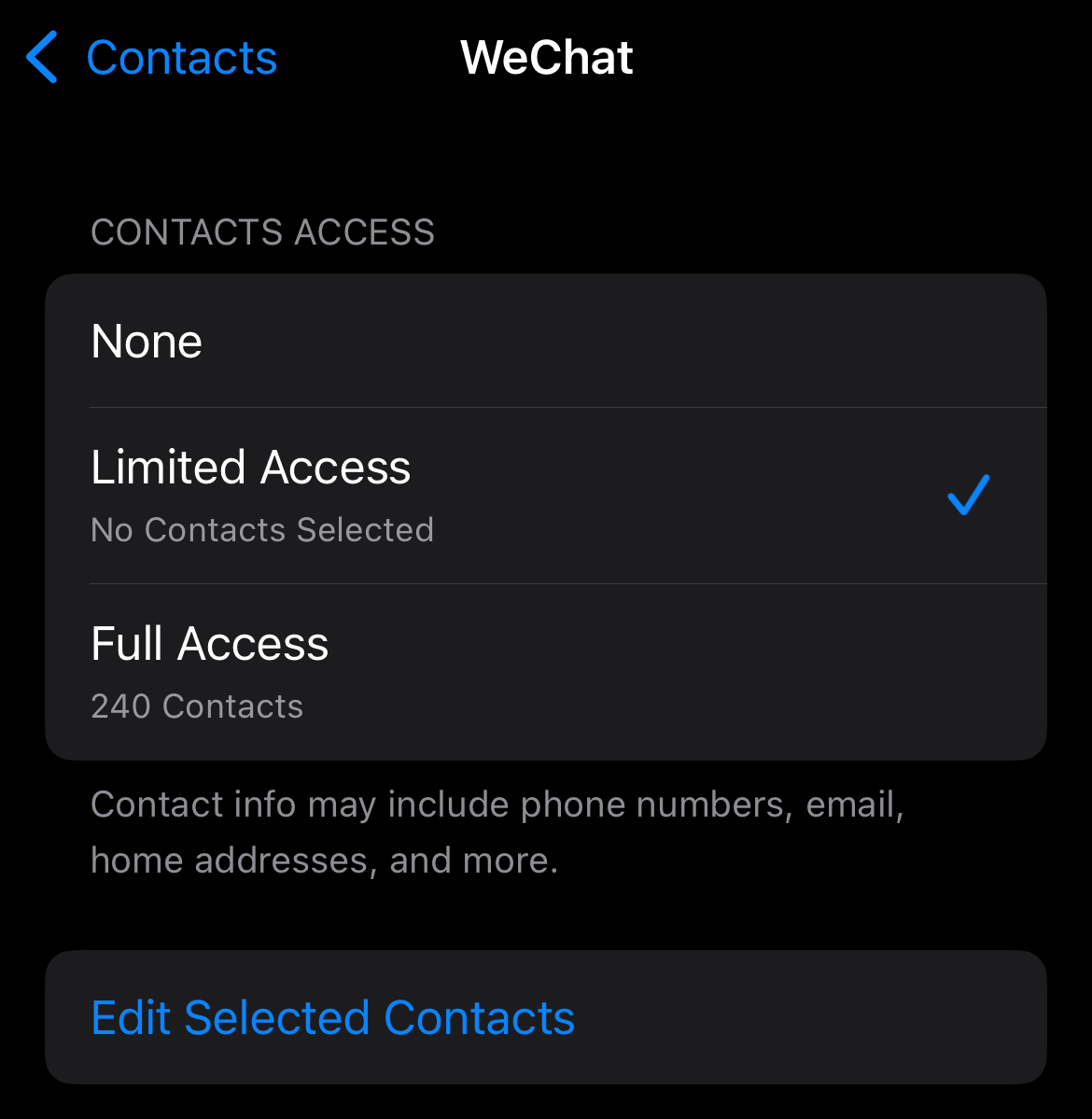}
        \caption{New OS feature: limited access to contacts}
        \label{fig:sys contact}
\end{figure}

For UU, WeChat does not provide tools to help users protect the privacy of others as much as possible while enjoying this function. However, we found that the latest operating system has made contributions in this regard. iOS provides countermeasure 2 for protecting the privacy of others as shown in Fig.~\ref{fig:sys contact}. 

\begin{figure}[tb]
        \centering\includegraphics[width=0.3\textwidth]{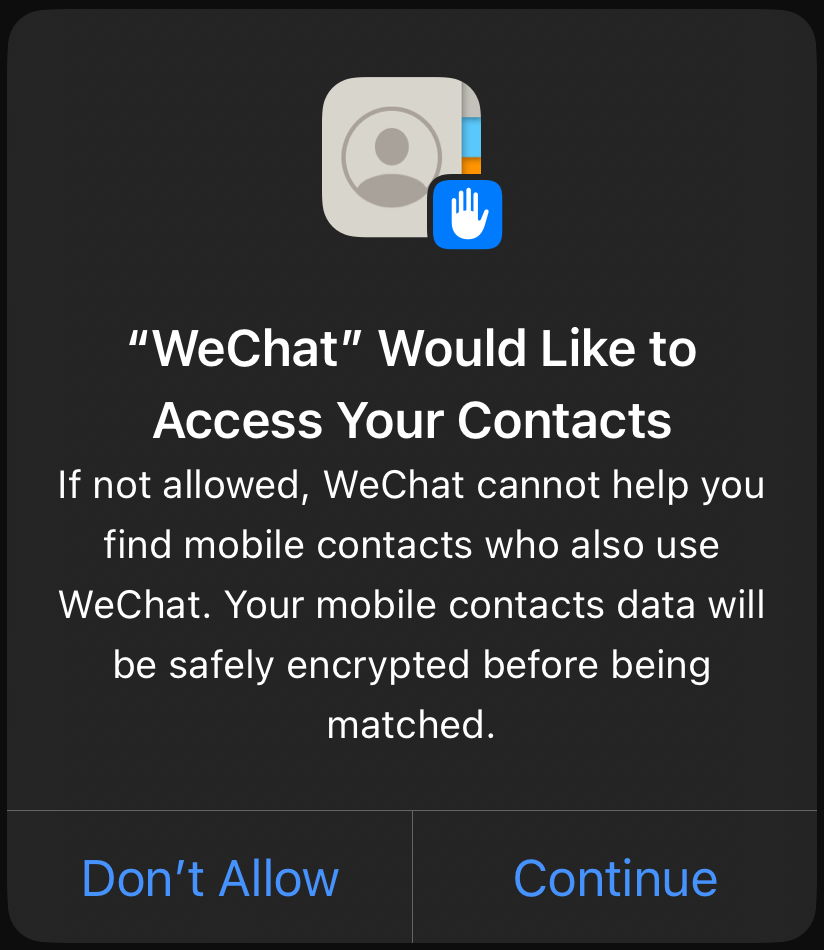}
        \caption{WeChat encrypts the address book}
        \label{fig:WeChat encrypts}
\end{figure}

Users can only partially upload the address book data. When UU clearly knows that some other users do not want others to upload their personal information, this corresponding measure can be used to protect the personal information of others. From the 6A analysis of the system, we know that awareness and authorization are clear for the notification and authorization of the data sender. WeChat will notify and obtain authorization in the form of a pop-up window (countermeasure 3), but it is currently impossible to notify and obtain authorization for users involved in the address book. Next, WeChat has taken measures to encrypt the address book information and ensure that the function of finding friends is realized by matching after encryption (countermeasure 4), as shown in Fig.~\ref{fig:WeChat encrypts}. The guarantee (countermeasure 5) derived from measure 4 and privacy policy (countermeasure 6) shows that if the original address book information is leaked by WeChat, the responsibility should be borne by WeChat. This combination of protection and accountability better protects user privacy, despite placing a greater burden on the system. Let us illustrate it with an example from the real world. Calendar applications are very common. When user A makes the meeting time and place of another user B public, the application has no right to analyze and process this information, and the application is not responsible for the resulting IDP information leakage. However, when the application provides an encryption function and the user chooses to encrypt user B's information before making the schedule public, the consequences of the information leakage caused by the application should be borne entirely by the application.

\noindent\textbf{WeChat payment: payee identity verification. }
In Fig.~\ref{fig:wechatDFD}, data flow $(3) \rightarrow (2)$, a specific area of concern is the WeChat money transfer feature. Unlike traditional bank transfers, WeChat allows users to attempt transfers without having full knowledge of the recipient's information. To prevent fraudulent transactions, WeChat employs a strategy where the system partially displays the recipient's name during the transfer process; the sender must correctly input the obscured portion of the name to complete the transaction. This approach, while intended as a security measure, inadvertently poses an IDP threat by allowing the sender to access partial name information of the recipient without their explicit consent. This not only compromises the privacy of the recipient but also undermines the confidentiality of personal data, as acquiring and confirming someone's name can be done with minimal effort and without a direct need to know that information for legitimate purposes. Although WeChat explained the alignment, saying that this practice complies with Chinese law\footnote{\url{https://www.gov.cn/zhengce/content/202312/content_6920724.htm}}, we believe there should be a better way.

\clearpage
\onecolumn
\begin{sidewaystable*}
    
\centering
\fontsize{6pt}{9pt}\selectfont
\addtolength{\tabcolsep}{-5pt}
\begin{longtable}[!h]{|l|l|l|l|l|l|l|l|l|l|l|l|l|l|l|}
\caption{Privacy threat analysis with IDPA}
\label{table:map-table}\\
\hline 
\multicolumn{6}{|p{383.375769pt}|}{\centering {\bfseries Interaction}} & \multicolumn{9}{|p{273.519231pt}|}{\centering {\bfseries Misactors \& Privacy Threat}}\\
\hline 
\multicolumn{15}{|p{656.895pt}|}{\centering {\bfseries User registration process and access management}}\\ 
\hline 
\multicolumn{1}{|p{100.33pt}}{\raggedright Source} & \multicolumn{1}{|p{100.44pt}}{\raggedright Flow} & \multicolumn{1}{|p{27.3519231pt}}{\raggedright IDPF} & \multicolumn{1}{|p{27.3519231pt}}{\raggedright PIDPF} & \multicolumn{1}{|p{27.3519231pt}}{\raggedright NIDPF}  &\multicolumn{1}{|p{100.55pt}}{\raggedright Destination} & 
\multicolumn{1}{|p{27.3519231pt}}{\raggedright MU} &
\multicolumn{1}{|p{27.3519231pt}}{\raggedright IU} & 
\multicolumn{1}{|p{27.3519231pt}}{\raggedright UU} & 
\multicolumn{1}{|p{27.3519231pt}}{\raggedright UFU} & 
\multicolumn{1}{|p{27.3519231pt}}{\raggedright SP} & 
\multicolumn{1}{|p{27.3519231pt}}{\raggedright GA} & 
\multicolumn{1}{|p{27.3519231pt}}{\raggedright IS} & 
\multicolumn{1}{|p{27.3519231pt}}{\raggedright IST} & 
\multicolumn{1}{|p{27.3519231pt}|}{\raggedright IP}\\ 
\hline 
\multicolumn{1}{|p{100.33pt}}{\raggedright Portal (2)} & \multicolumn{1}{|p{100.44pt}}{\raggedright Registration request} & \multicolumn{1}{|p{27.3519231pt}}{\raggedright } & \multicolumn{1}{|p{27.3519231pt}}{\raggedright } & \multicolumn{1}{|p{27.3519231pt}}{\centering x} & \multicolumn{1}{|p{100.55pt}}{\raggedright Registration Service (11)} & \multicolumn{1}{|p{27.3519231pt}}{} & \multicolumn{1}{|p{27.3519231pt}}{\centering }  & \multicolumn{1}{|p{27.3519231pt}}{} & \multicolumn{1}{|p{27.3519231pt}}{} & \multicolumn{1}{|p{27.3519231pt}}{} & \multicolumn{1}{|p{27.3519231pt}}{} & \multicolumn{1}{|p{27.3519231pt}}{} & \multicolumn{1}{|p{27.3519231pt}}{}  & \multicolumn{1}{|p{27.3519231pt}|}{} \\
\hline
\multicolumn{1}{|p{100.33pt}}{\raggedright Registration Service (11) } & \multicolumn{1}{|p{100.44pt}}{\raggedright Registration response} & \multicolumn{1}{|p{27.3519231pt}}{\centering } & \multicolumn{1}{|p{27.3519231pt}}{\centering x} & \multicolumn{1}{|p{27.3519231pt}}{} & \multicolumn{1}{|p{100.55pt}}{\raggedright Portal (2)} & \multicolumn{1}{|p{27.3519231pt}}{\centering x} & \multicolumn{1}{|p{27.3519231pt}}{\centering x}  & \multicolumn{1}{|p{27.3519231pt}}{} & \multicolumn{1}{|p{27.3519231pt}}{\centering x} & \multicolumn{1}{|p{27.3519231pt}}{\centering x} & \multicolumn{1}{|p{27.3519231pt}}{} & \multicolumn{1}{|p{27.3519231pt}}{\centering x} & \multicolumn{1}{|p{27.3519231pt}}{}  & \multicolumn{1}{|p{27.3519231pt}|}{} \\ 
\hline 
\multicolumn{1}{|p{100.33pt}}{\raggedright Registration service (11)} & \multicolumn{1}{|p{100.44pt}}{\raggedright Request} & \multicolumn{1}{|p{27.3519231pt}}{\raggedright } & \multicolumn{1}{|p{27.3519231pt}}{\raggedright } & \multicolumn{1}{|p{27.3519231pt}}{\centering x} &\multicolumn{1}{|p{100.55pt}}{\raggedright User Profile (7)} & \multicolumn{1}{|p{27.3519231pt}}{} & \multicolumn{1}{|p{27.3519231pt}}{} & \multicolumn{1}{|p{27.3519231pt}}{} & \multicolumn{1}{|p{27.3519231pt}}{} & \multicolumn{1}{|p{27.3519231pt}}{\centering } & \multicolumn{1}{|p{27.3519231pt}}{} & \multicolumn{1}{|p{27.3519231pt}}{} & \multicolumn{1}{|p{27.3519231pt}}{} & \multicolumn{1}{|p{27.3519231pt}|}{\centering } \\ 
 \hline 
\multicolumn{1}{|p{100.33pt}}{\raggedright User Profile (7)} & \multicolumn{1}{|p{100.44pt}}{\raggedright Response} & \multicolumn{1}{|p{27.3519231pt}}{\raggedright } & \multicolumn{1}{|p{27.3519231pt}}{\centering x } & \multicolumn{1}{|p{27.3519231pt}}{\centering } &\multicolumn{1}{|p{100.55pt}}{\raggedright Registration service (11)} & \multicolumn{1}{|p{27.3519231pt}}{\centering x} & \multicolumn{1}{|p{27.3519231pt}}{\centering x} & \multicolumn{1}{|p{27.3519231pt}}{} & \multicolumn{1}{|p{27.3519231pt}}{\centering x} & \multicolumn{1}{|p{27.3519231pt}}{\centering x} & \multicolumn{1}{|p{27.3519231pt}}{} & \multicolumn{1}{|p{27.3519231pt}}{\centering x} & \multicolumn{1}{|p{27.3519231pt}}{} & \multicolumn{1}{|p{27.3519231pt}|}{\centering } \\ 
 \hline 
\multicolumn{1}{|p{100.33pt}}{\raggedright Registration Service (11)} & \multicolumn{1}{|p{100.44pt}}{\raggedright Request} & \multicolumn{1}{|p{27.3519231pt}}{\raggedright } & \multicolumn{1}{|p{27.3519231pt}}{\raggedright } & \multicolumn{1}{|p{27.3519231pt}}{\centering x} &\multicolumn{1}{|p{100.55pt}}{\raggedright Login Process (12)} & \multicolumn{1}{|p{27.3519231pt}}{} & \multicolumn{1}{|p{27.3519231pt}}{} & \multicolumn{1}{|p{27.3519231pt}}{} & \multicolumn{1}{|p{27.3519231pt}}{} & \multicolumn{1}{|p{27.3519231pt}}{\centering } & \multicolumn{1}{|p{27.3519231pt}}{} & \multicolumn{1}{|p{27.3519231pt}}{} & \multicolumn{1}{|p{27.3519231pt}}{} & \multicolumn{1}{|p{27.3519231pt}|}{\centering } \\ 
 \hline 
\multicolumn{1}{|p{100.33pt}}{\raggedright Login Process (12)} & \multicolumn{1}{|p{100.44pt}}{\raggedright Response} & \multicolumn{1}{|p{27.3519231pt}}{\raggedright } & \multicolumn{1}{|p{27.3519231pt}}{\raggedright } & \multicolumn{1}{|p{27.3519231pt}}{\centering x} &\multicolumn{1}{|p{100.55pt}}{\raggedright Registration Service (11)} & \multicolumn{1}{|p{27.3519231pt}}{} & \multicolumn{1}{|p{27.3519231pt}}{} & \multicolumn{1}{|p{27.3519231pt}}{} & \multicolumn{1}{|p{27.3519231pt}}{} & \multicolumn{1}{|p{27.3519231pt}}{\centering } & \multicolumn{1}{|p{27.3519231pt}}{} & \multicolumn{1}{|p{27.3519231pt}}{} & \multicolumn{1}{|p{27.3519231pt}}{} & \multicolumn{1}{|p{27.3519231pt}|}{\centering } \\ 
 \hline 
\multicolumn{1}{|p{100.33pt}}{\raggedright Portal (2)} & \multicolumn{1}{|p{100.44pt}}{\raggedright Request} & \multicolumn{1}{|p{27.3519231pt}}{\raggedright } & \multicolumn{1}{|p{27.3519231pt}}{\raggedright } & \multicolumn{1}{|p{27.3519231pt}}{\centering x} &\multicolumn{1}{|p{100.55pt}}{\raggedright Login Process (12)} & \multicolumn{1}{|p{27.3519231pt}}{} & \multicolumn{1}{|p{27.3519231pt}}{} & \multicolumn{1}{|p{27.3519231pt}}{} & \multicolumn{1}{|p{27.3519231pt}}{} & \multicolumn{1}{|p{27.3519231pt}}{\centering } & \multicolumn{1}{|p{27.3519231pt}}{} & \multicolumn{1}{|p{27.3519231pt}}{} & \multicolumn{1}{|p{27.3519231pt}}{} & \multicolumn{1}{|p{27.3519231pt}|}{\centering } \\ 
 \hline 
\multicolumn{1}{|p{100.33pt}}{\raggedright Login Process (12)} & \multicolumn{1}{|p{100.44pt}}{\raggedright Response} & \multicolumn{1}{|p{27.3519231pt}}{\raggedright } & \multicolumn{1}{|p{27.3519231pt}}{\raggedright } & \multicolumn{1}{|p{27.3519231pt}}{\centering x} &\multicolumn{1}{|p{100.55pt}}{\raggedright Portal (2)} & \multicolumn{1}{|p{27.3519231pt}}{} & \multicolumn{1}{|p{27.3519231pt}}{} & \multicolumn{1}{|p{27.3519231pt}}{} & \multicolumn{1}{|p{27.3519231pt}}{} & \multicolumn{1}{|p{27.3519231pt}}{\centering } & \multicolumn{1}{|p{27.3519231pt}}{} & \multicolumn{1}{|p{27.3519231pt}}{} & \multicolumn{1}{|p{27.3519231pt}}{} & \multicolumn{1}{|p{27.3519231pt}|}{\centering } \\ 
 \hline 
\multicolumn{1}{|p{100.33pt}}{\raggedright Login Process (12)} & \multicolumn{1}{|p{100.44pt}}{\raggedright Request} & \multicolumn{1}{|p{27.3519231pt}}{\raggedright } & \multicolumn{1}{|p{27.3519231pt}}{\raggedright } & \multicolumn{1}{|p{27.3519231pt}}{\centering x} &\multicolumn{1}{|p{100.55pt}}{\raggedright Authentication Process (13)} & \multicolumn{1}{|p{27.3519231pt}}{} & \multicolumn{1}{|p{27.3519231pt}}{} & \multicolumn{1}{|p{27.3519231pt}}{} & \multicolumn{1}{|p{27.3519231pt}}{} & \multicolumn{1}{|p{27.3519231pt}}{\centering } & \multicolumn{1}{|p{27.3519231pt}}{} & \multicolumn{1}{|p{27.3519231pt}}{} & \multicolumn{1}{|p{27.3519231pt}}{} & \multicolumn{1}{|p{27.3519231pt}|}{\centering } \\ 
 \hline 
\multicolumn{1}{|p{100.33pt}}{\raggedright Authentication Process (13)} & \multicolumn{1}{|p{100.44pt}}{\raggedright Response} & \multicolumn{1}{|p{27.3519231pt}}{\raggedright } & \multicolumn{1}{|p{27.3519231pt}}{\raggedright } & \multicolumn{1}{|p{27.3519231pt}}{\centering x} &\multicolumn{1}{|p{100.55pt}}{\raggedright Login Process (12)} & \multicolumn{1}{|p{27.3519231pt}}{} & \multicolumn{1}{|p{27.3519231pt}}{} & \multicolumn{1}{|p{27.3519231pt}}{} & \multicolumn{1}{|p{27.3519231pt}}{} & \multicolumn{1}{|p{27.3519231pt}}{\centering } & \multicolumn{1}{|p{27.3519231pt}}{} & \multicolumn{1}{|p{27.3519231pt}}{} & \multicolumn{1}{|p{27.3519231pt}}{} & \multicolumn{1}{|p{27.3519231pt}|}{\centering } \\ 
 \hline 
\multicolumn{1}{|p{100.33pt}}{\raggedright Authentication Process (13)} & \multicolumn{1}{|p{100.44pt}}{\raggedright Request} & \multicolumn{1}{|p{27.3519231pt}}{\raggedright } & \multicolumn{1}{|p{27.3519231pt}}{\raggedright } & \multicolumn{1}{|p{27.3519231pt}}{\centering x} &\multicolumn{1}{|p{100.55pt}}{\raggedright Identity and Authentication Manager (14)} & \multicolumn{1}{|p{27.3519231pt}}{} & \multicolumn{1}{|p{27.3519231pt}}{} & \multicolumn{1}{|p{27.3519231pt}}{} & \multicolumn{1}{|p{27.3519231pt}}{} & \multicolumn{1}{|p{27.3519231pt}}{\centering } & \multicolumn{1}{|p{27.3519231pt}}{} & \multicolumn{1}{|p{27.3519231pt}}{} & \multicolumn{1}{|p{27.3519231pt}}{} & \multicolumn{1}{|p{27.3519231pt}|}{\centering } \\ 
 \hline 
\multicolumn{1}{|p{100.33pt}}{\raggedright Identity and Authentication Manager (14)} & \multicolumn{1}{|p{100.44pt}}{\raggedright Response} & \multicolumn{1}{|p{27.3519231pt}}{\raggedright } & \multicolumn{1}{|p{27.3519231pt}}{\raggedright } & \multicolumn{1}{|p{27.3519231pt}}{\centering x} &\multicolumn{1}{|p{100.55pt}}{\raggedright Authentication Process (13)} & \multicolumn{1}{|p{27.3519231pt}}{} & \multicolumn{1}{|p{27.3519231pt}}{} & \multicolumn{1}{|p{27.3519231pt}}{} & \multicolumn{1}{|p{27.3519231pt}}{} & \multicolumn{1}{|p{27.3519231pt}}{\centering } & \multicolumn{1}{|p{27.3519231pt}}{} & \multicolumn{1}{|p{27.3519231pt}}{} & \multicolumn{1}{|p{27.3519231pt}}{} & \multicolumn{1}{|p{27.3519231pt}|}{\centering } \\ 
 \hline 
\multicolumn{1}{|p{100.33pt}}{\raggedright Identity and Authentication Manager (14)} & \multicolumn{1}{|p{100.44pt}}{\raggedright Request} & \multicolumn{1}{|p{27.3519231pt}}{\raggedright } & \multicolumn{1}{|p{27.3519231pt}}{\raggedright } & \multicolumn{1}{|p{27.3519231pt}}{\centering x} &\multicolumn{1}{|p{100.55pt}}{\raggedright Data Access Setting (15)} & \multicolumn{1}{|p{27.3519231pt}}{} & \multicolumn{1}{|p{27.3519231pt}}{} & \multicolumn{1}{|p{27.3519231pt}}{} & \multicolumn{1}{|p{27.3519231pt}}{} & \multicolumn{1}{|p{27.3519231pt}}{\centering } & \multicolumn{1}{|p{27.3519231pt}}{} & \multicolumn{1}{|p{27.3519231pt}}{} & \multicolumn{1}{|p{27.3519231pt}}{} & \multicolumn{1}{|p{27.3519231pt}|}{\centering } \\ 
 \hline 
\multicolumn{1}{|p{100.33pt}}{\raggedright Data Access Setting (15)} & \multicolumn{1}{|p{100.44pt}}{\raggedright Response} & \multicolumn{1}{|p{27.3519231pt}}{\raggedright } & \multicolumn{1}{|p{27.3519231pt}}{\raggedright } & \multicolumn{1}{|p{27.3519231pt}}{\centering x} &\multicolumn{1}{|p{100.55pt}}{\raggedright Identity and Authentication Manager (14)} & \multicolumn{1}{|p{27.3519231pt}}{} & \multicolumn{1}{|p{27.3519231pt}}{} & \multicolumn{1}{|p{27.3519231pt}}{} & \multicolumn{1}{|p{27.3519231pt}}{} & \multicolumn{1}{|p{27.3519231pt}}{\centering } & \multicolumn{1}{|p{27.3519231pt}}{} & \multicolumn{1}{|p{27.3519231pt}}{} & \multicolumn{1}{|p{27.3519231pt}}{} & \multicolumn{1}{|p{27.3519231pt}|}{\centering } \\ 
 \hline  
\multicolumn{1}{|p{100.33pt}}{\raggedright User Profile (7)} & \multicolumn{1}{|p{100.44pt}}{\raggedright Data} & \multicolumn{1}{|p{27.3519231pt}}{\raggedright } & \multicolumn{1}{|p{27.3519231pt}}{\raggedright } & \multicolumn{1}{|p{27.3519231pt}}{\centering x} &\multicolumn{1}{|p{100.55pt}}{\raggedright Data Access Setting (15)} & \multicolumn{1}{|p{27.3519231pt}}{} & \multicolumn{1}{|p{27.3519231pt}}{} & \multicolumn{1}{|p{27.3519231pt}}{} & \multicolumn{1}{|p{27.3519231pt}}{} & \multicolumn{1}{|p{27.3519231pt}}{\centering } & \multicolumn{1}{|p{27.3519231pt}}{} & \multicolumn{1}{|p{27.3519231pt}}{} & \multicolumn{1}{|p{27.3519231pt}}{} & \multicolumn{1}{|p{27.3519231pt}|}{\centering } \\ 
 \hline   
\multicolumn{15}{|p{627.7825pt}|}{\centering {\bfseries Third-party access}}\\ 
 \hline 
\multicolumn{1}{|p{100.33pt}}{\raggedright Source/Destination} & \multicolumn{1}{|p{100.44pt}}{\raggedright Flow} & \multicolumn{1}{|p{27.3519231pt}}{\raggedright IDPF} & \multicolumn{1}{|p{27.3519231pt}}{\raggedright PIDPF} & \multicolumn{1}{|p{27.3519231pt}}{\raggedright NIDPF} &\multicolumn{1}{|p{100.55pt}}{\raggedright Destination/Source} & \multicolumn{1}{|p{27.3519231pt}}{\raggedright MU} & \multicolumn{1}{|p{27.3519231pt}}{\raggedright IU} & \multicolumn{1}{|p{27.3519231pt}}{\raggedright UU} & \multicolumn{1}{|p{27.3519231pt}}{\raggedright UFU} & \multicolumn{1}
{|p{27.3519231pt}}{\raggedright SP} & \multicolumn{1}
{|p{27.3519231pt}}{\raggedright GA} & \multicolumn{1}{|p{27.3519231pt}}{\raggedright IS} & \multicolumn{1}{|p{27.3519231pt}}{\raggedright IST} & \multicolumn{1}{|p{27.3519231pt}|}{\raggedright IP}\\ 
 \hline 
\multicolumn{1}{|p{100.33pt}}{\raggedright Third-party (16)} & \multicolumn{1}{|p{100.44pt}}{\raggedright Access request/response} & \multicolumn{1}{|p{27.3519231pt}}{\centering } & \multicolumn{1}{|p{27.3519231pt}}{\centering x} & \multicolumn{1}{|p{27.3519231pt}}{\centering } &\multicolumn{1}{|p{100.55pt}}{\raggedright Data Access Setting (15)} & \multicolumn{1}{|p{27.3519231pt}}{\centering x} & \multicolumn{1}{|p{27.3519231pt}}{\centering x} & \multicolumn{1}{|p{27.3519231pt}}{\centering x} & \multicolumn{1}{|p{27.3519231pt}}{\centering x} & \multicolumn{1}{|p{27.3519231pt}}{\centering x} & \multicolumn{1}
{|p{27.3519231pt}}{} & \multicolumn{1}{|p{27.3519231pt}}{\centering x} & \multicolumn{1}{|p{27.3519231pt}}{\centering x} & \multicolumn{1}{|p{27.3519231pt}|}{\centering x}\\ 
\hline
\multicolumn{15}{|p{627.7825pt}|}{\centering {\bfseries Application Service Processing}}\\ 
\hline 
\multicolumn{1}{|p{100.33pt}}{\raggedright Source} & \multicolumn{1}{|p{100.44pt}}{\raggedright Flow} & \multicolumn{1}{|p{27.3519231pt}}{\raggedright IDPF} & \multicolumn{1}{|p{27.3519231pt}}{\raggedright PIDPF} & \multicolumn{1}{|p{27.3519231pt}}{\raggedright NIDPF} &\multicolumn{1}{|p{100.55pt}}{\raggedright Destination} & \multicolumn{1}{|p{27.3519231pt}}{\raggedright MU} & \multicolumn{1}{|p{27.3519231pt}}{\raggedright IU} & \multicolumn{1}{|p{27.3519231pt}}{\raggedright UU} & \multicolumn{1}{|p{27.3519231pt}}{\raggedright UFU} & \multicolumn{1}
{|p{27.3519231pt}}{\raggedright SP} & \multicolumn{1}{|p{27.3519231pt}}{\raggedright GA} & \multicolumn{1}{|p{27.3519231pt}}{\raggedright IS} & \multicolumn{1}{|p{27.3519231pt}}{\raggedright IST} & \multicolumn{1}{|p{27.3519231pt}|}{\raggedright IP}\\ 
\hline 
\multicolumn{1}{|p{100.33pt}}{\raggedright Portal (2)} & \multicolumn{1}{|p{100.44pt}}{\raggedright Transfer request} & \multicolumn{1}{|p{27.3519231pt}}{\centering } & \multicolumn{1}{|p{27.3519231pt}}{\centering } & \multicolumn{1}{|p{27.3519231pt}}{\centering x} &\multicolumn{1}{|p{100.55pt}}{\raggedright Payment Processing (3)} & \multicolumn{1}{|p{27.3519231pt}}{\centering } & \multicolumn{1}{|p{27.3519231pt}}{\centering } & \multicolumn{1}{|p{27.3519231pt}}{\centering } & \multicolumn{1}{|p{27.3519231pt}}{\centering } & \multicolumn{1}{|p{27.3519231pt}}{} & \multicolumn{1}{|p{27.3519231pt}}{\centering } & \multicolumn{1}{|p{27.3519231pt}}{} & \multicolumn{1}{|p{27.3519231pt}}{} & \multicolumn{1}{|p{27.3519231pt}|}{}\\ 
\hline
\multicolumn{1}{|p{100.33pt}}{\raggedright Payment Processing (3)} & \multicolumn{1}{|p{100.44pt}}{\raggedright Recipient verication} & \multicolumn{1}{|p{27.3519231pt}}{\centering x} & \multicolumn{1}{|p{27.3519231pt}}{\centering } & \multicolumn{1}{|p{27.3519231pt}}{\centering } &\multicolumn{1}{|p{100.55pt}}{\raggedright Portal (2)} & \multicolumn{1}{|p{27.3519231pt}}{\centering x} & \multicolumn{1}{|p{27.3519231pt}}{\centering x} & \multicolumn{1}{|p{27.3519231pt}}{\centering x} & \multicolumn{1}{|p{27.3519231pt}}{\centering x} & \multicolumn{1}{|p{27.3519231pt}}{\centering x} & \multicolumn{1}{|p{27.3519231pt}}{\centering } & \multicolumn{1}{|p{27.3519231pt}}{\centering x} & \multicolumn{1}{|p{27.3519231pt}}{} & \multicolumn{1}{|p{27.3519231pt}|}{}\\ 
\hline 
\multicolumn{1}{|p{100.33pt}}{\raggedright Payment Processing (3)} & \multicolumn{1}{|p{100.44pt}}{\raggedright Data retrieval} & \multicolumn{1}{|p{27.3519231pt}}{\centering } & \multicolumn{1}{|p{27.3519231pt}}{\centering } & \multicolumn{1}{|p{27.3519231pt}}{\centering x} &\multicolumn{1}{|p{100.55pt}}{\raggedright User Profile (7)} & \multicolumn{1}{|p{27.3519231pt}}{\centering } & \multicolumn{1}{|p{27.3519231pt}}{\centering } & \multicolumn{1}{|p{27.3519231pt}}{\centering } & \multicolumn{1}{|p{27.3519231pt}}{\centering } & \multicolumn{1}{|p{27.3519231pt}}{\centering } & \multicolumn{1}{|p{27.3519231pt}}{\centering } & \multicolumn{1}{|p{27.3519231pt}}{\centering } & \multicolumn{1}{|p{27.3519231pt}}{} & \multicolumn{1}{|p{27.3519231pt}|}{}\\ 
\hline 
\multicolumn{1}{|p{100.33pt}}{\raggedright User Profile (7)} & \multicolumn{1}{|p{100.44pt}}{\raggedright Recipient data} & \multicolumn{1}{|p{27.3519231pt}}{\centering x} & \multicolumn{1}{|p{27.3519231pt}}{\centering } & \multicolumn{1}{|p{27.3519231pt}}{\centering } &\multicolumn{1}{|p{100.55pt}}{\raggedright Payment Processing (2)} & \multicolumn{1}{|p{27.3519231pt}}{\centering x} & \multicolumn{1}{|p{27.3519231pt}}{\centering x} & \multicolumn{1}{|p{27.3519231pt}}{\centering x} & \multicolumn{1}{|p{27.3519231pt}}{\centering x} & \multicolumn{1}{|p{27.3519231pt}}{\centering x} & \multicolumn{1}{|p{27.3519231pt}}{\centering } & \multicolumn{1}{|p{27.3519231pt}}{\centering x} & \multicolumn{1}{|p{27.3519231pt}}{\centering } & \multicolumn{1}{|p{27.3519231pt}|}{\centering x}\\ 
\hline
\multicolumn{1}{|p{100.33pt}}{\raggedright Portal (2)} & \multicolumn{1}{|p{100.44pt}}{\raggedright Post data} & \multicolumn{1}{|p{27.3519231pt}}{\centering } & \multicolumn{1}{|p{27.3519231pt}}{\centering x} & \multicolumn{1}{|p{27.3519231pt}}{\centering } &\multicolumn{1}{|p{100.55pt}}{\raggedright Social Media Functions (4)} & \multicolumn{1}{|p{27.3519231pt}}{\centering x} & \multicolumn{1}{|p{27.3519231pt}}{\centering x} & \multicolumn{1}{|p{27.3519231pt}}{\centering x} & \multicolumn{1}{|p{27.3519231pt}}{\centering x} & \multicolumn{1}{|p{27.3519231pt}}{\centering x} & \multicolumn{1}{|p{27.3519231pt}}{\centering } & \multicolumn{1}{|p{27.3519231pt}}{\centering x} & \multicolumn{1}{|p{27.3519231pt}}{\centering } & \multicolumn{1}{|p{27.3519231pt}|}{\centering x}\\ 
\hline
\multicolumn{1}{|p{100.33pt}}{\raggedright Social Media Functions (4)} & \multicolumn{1}{|p{100.44pt}}{\raggedright Post data} & \multicolumn{1}{|p{27.3519231pt}}{\centering } & \multicolumn{1}{|p{27.3519231pt}}{\centering x} & \multicolumn{1}{|p{27.3519231pt}}{\centering } &\multicolumn{1}{|p{100.55pt}}{\raggedright Portal (2)} & \multicolumn{1}{|p{27.3519231pt}}{\centering x} & \multicolumn{1}{|p{27.3519231pt}}{\centering x} & \multicolumn{1}{|p{27.3519231pt}}{\centering x} & \multicolumn{1}{|p{27.3519231pt}}{\centering x} & \multicolumn{1}{|p{27.3519231pt}}{\centering x} & \multicolumn{1}{|p{27.3519231pt}}{\centering } & \multicolumn{1}{|p{27.3519231pt}}{\centering x} & \multicolumn{1}{|p{27.3519231pt}}{\centering } & \multicolumn{1}{|p{27.3519231pt}|}{\centering x}\\ 
\hline
\multicolumn{1}{|p{100.33pt}}{\raggedright Social Media Functions (4)} & \multicolumn{1}{|p{100.44pt}}{\raggedright Post data} & \multicolumn{1}{|p{27.3519231pt}}{\centering } & \multicolumn{1}{|p{27.3519231pt}}{\centering x} & \multicolumn{1}{|p{27.3519231pt}}{\centering } &\multicolumn{1}{|p{100.55pt}}{\raggedright Post Data (8)} & \multicolumn{1}{|p{27.3519231pt}}{\centering x} & \multicolumn{1}{|p{27.3519231pt}}{\centering x} & \multicolumn{1}{|p{27.3519231pt}}{\centering x} & \multicolumn{1}{|p{27.3519231pt}}{\centering x} & \multicolumn{1}{|p{27.3519231pt}}{\centering x} & \multicolumn{1}{|p{27.3519231pt}}{\centering } & \multicolumn{1}{|p{27.3519231pt}}{\centering } & \multicolumn{1}{|p{27.3519231pt}}{\centering x} & \multicolumn{1}{|p{27.3519231pt}|}{\centering x}\\ 
\hline
\multicolumn{1}{|p{100.33pt}}{\raggedright Post Data (8)} & \multicolumn{1}{|p{100.44pt}}{\raggedright Post data} & \multicolumn{1}{|p{27.3519231pt}}{\centering } & \multicolumn{1}{|p{27.3519231pt}}{\centering x} & \multicolumn{1}{|p{27.3519231pt}}{\centering } &\multicolumn{1}{|p{100.55pt}}{\raggedright Social Media Function (4)} & \multicolumn{1}{|p{27.3519231pt}}{\centering x} & \multicolumn{1}{|p{27.3519231pt}}{\centering x} & \multicolumn{1}{|p{27.3519231pt}}{\centering x} & \multicolumn{1}{|p{27.3519231pt}}{\centering x} & \multicolumn{1}{|p{27.3519231pt}}{\centering x} & \multicolumn{1}{|p{27.3519231pt}}{\centering } & \multicolumn{1}{|p{27.3519231pt}}{\centering x} & \multicolumn{1}{|p{27.3519231pt}}{\centering } & \multicolumn{1}{|p{27.3519231pt}|}{\centering x}\\ 
\hline
\multicolumn{1}{|p{100.33pt}}{\raggedright Portal (2)} & \multicolumn{1}{|p{100.44pt}}{\raggedright Message data} & \multicolumn{1}{|p{27.3519231pt}}{\centering } & \multicolumn{1}{|p{27.3519231pt}}{\centering x} & \multicolumn{1}{|p{27.3519231pt}}{\centering } &\multicolumn{1}{|p{100.55pt}}{\raggedright Messaging (5)} & \multicolumn{1}{|p{27.3519231pt}}{\centering x} & \multicolumn{1}{|p{27.3519231pt}}{\centering x} & \multicolumn{1}{|p{27.3519231pt}}{\centering x} & \multicolumn{1}{|p{27.3519231pt}}{\centering x} & \multicolumn{1}{|p{27.3519231pt}}{\centering x} & \multicolumn{1}{|p{27.3519231pt}}{\centering } & \multicolumn{1}{|p{27.3519231pt}}{\centering x} & \multicolumn{1}{|p{27.3519231pt}}{\centering x} & \multicolumn{1}{|p{27.3519231pt}|}{\centering }\\ 
\hline
\multicolumn{1}{|p{100.33pt}}{\raggedright Messaging (5)} & \multicolumn{1}{|p{100.44pt}}{\raggedright Message data} & \multicolumn{1}{|p{27.3519231pt}}{\centering } & \multicolumn{1}{|p{27.3519231pt}}{\centering x} & \multicolumn{1}{|p{27.3519231pt}}{\centering } &\multicolumn{1}{|p{100.55pt}}{\raggedright Portal (2)} & \multicolumn{1}{|p{27.3519231pt}}{\centering x} & \multicolumn{1}{|p{27.3519231pt}}{\centering x} & \multicolumn{1}{|p{27.3519231pt}}{\centering x} & \multicolumn{1}{|p{27.3519231pt}}{\centering x} & \multicolumn{1}{|p{27.3519231pt}}{\centering x} & \multicolumn{1}{|p{27.3519231pt}}{\centering } & \multicolumn{1}{|p{27.3519231pt}}{\centering x} & \multicolumn{1}{|p{27.3519231pt}}{\centering x} & \multicolumn{1}{|p{27.3519231pt}|}{\centering }\\ 
\hline
\multicolumn{1}{|p{100.33pt}}{\raggedright Messaging (5)} & \multicolumn{1}{|p{100.44pt}}{\raggedright Message data} & \multicolumn{1}{|p{27.3519231pt}}{\centering } & \multicolumn{1}{|p{27.3519231pt}}{\centering x} & \multicolumn{1}{|p{27.3519231pt}}{\centering } &\multicolumn{1}{|p{100.55pt}}{\raggedright Message History (9)} & \multicolumn{1}{|p{27.3519231pt}}{\centering x} & \multicolumn{1}{|p{27.3519231pt}}{\centering x} & \multicolumn{1}{|p{27.3519231pt}}{\centering x} & \multicolumn{1}{|p{27.3519231pt}}{\centering x} & \multicolumn{1}{|p{27.3519231pt}}{\centering x} & \multicolumn{1}{|p{27.3519231pt}}{\centering } & \multicolumn{1}{|p{27.3519231pt}}{\centering } & \multicolumn{1}{|p{27.3519231pt}}{\centering x} & \multicolumn{1}{|p{27.3519231pt}|}{\centering x}\\ 
\hline
\multicolumn{1}{|p{100.33pt}}{\raggedright Message History (9)} & \multicolumn{1}{|p{100.44pt}}{\raggedright Message data} & \multicolumn{1}{|p{27.3519231pt}}{\centering } & \multicolumn{1}{|p{27.3519231pt}}{\centering x} & \multicolumn{1}{|p{27.3519231pt}}{\centering } &\multicolumn{1}{|p{100.55pt}}{\raggedright Messaging (5)} & \multicolumn{1}{|p{27.3519231pt}}{\centering x} & \multicolumn{1}{|p{27.3519231pt}}{\centering x} & \multicolumn{1}{|p{27.3519231pt}}{\centering x} & \multicolumn{1}{|p{27.3519231pt}}{\centering x} & \multicolumn{1}{|p{27.3519231pt}}{\centering x} & \multicolumn{1}{|p{27.3519231pt}}{\centering } & \multicolumn{1}{|p{27.3519231pt}}{\centering x} & \multicolumn{1}{|p{27.3519231pt}}{\centering } & \multicolumn{1}{|p{27.3519231pt}|}{\centering x}\\ 
\hline
\multicolumn{1}{|p{100.33pt}}{\raggedright Portal (2)} & \multicolumn{1}{|p{100.44pt}}{\raggedright Upload contact list} & \multicolumn{1}{|p{27.3519231pt}}{\centering x} & \multicolumn{1}{|p{27.3519231pt}}{\centering } & \multicolumn{1}{|p{27.3519231pt}}{\centering } &\multicolumn{1}{|p{100.55pt}}{\raggedright Find friends via contact list (6)} & \multicolumn{1}{|p{27.3519231pt}}{\centering x} & \multicolumn{1}{|p{27.3519231pt}}{\centering x} & \multicolumn{1}{|p{27.3519231pt}}{\centering x} & \multicolumn{1}{|p{27.3519231pt}}{\centering x} & \multicolumn{1}{|p{27.3519231pt}}{\centering x} & \multicolumn{1}{|p{27.3519231pt}}{\centering } & \multicolumn{1}{|p{27.3519231pt}}{\centering x} & \multicolumn{1}{|p{27.3519231pt}}{\centering } & \multicolumn{1}{|p{27.3519231pt}|}{\centering x}\\ 
\hline
\multicolumn{1}{|p{100.33pt}}{\raggedright Find friends via contact list (6)} & \multicolumn{1}{|p{100.44pt}}{\raggedright User Relationship} & \multicolumn{1}{|p{27.3519231pt}}{\centering x} & \multicolumn{1}{|p{27.3519231pt}}{\centering } & \multicolumn{1}{|p{27.3519231pt}}{\centering } &\multicolumn{1}{|p{100.55pt}}{\raggedright Portal (2)} & \multicolumn{1}{|p{27.3519231pt}}{\centering x} & \multicolumn{1}{|p{27.3519231pt}}{\centering x} & \multicolumn{1}{|p{27.3519231pt}}{\centering x} & \multicolumn{1}{|p{27.3519231pt}}{\centering x} & \multicolumn{1}{|p{27.3519231pt}}{\centering x} & \multicolumn{1}{|p{27.3519231pt}}{\centering } & \multicolumn{1}{|p{27.3519231pt}}{\centering x} & \multicolumn{1}{|p{27.3519231pt}}{\centering } & \multicolumn{1}{|p{27.3519231pt}|}{\centering x}\\ 
\hline
\multicolumn{1}{|p{100.33pt}}{\raggedright Find friends via contact list (6)} & \multicolumn{1}{|p{100.44pt}}{\raggedright User Relationship data} & \multicolumn{1}{|p{27.3519231pt}}{\centering x} & \multicolumn{1}{|p{27.3519231pt}}{\centering } & \multicolumn{1}{|p{27.3519231pt}}{\centering } &\multicolumn{1}{|p{100.55pt}}{\raggedright User Relationship (10)} & \multicolumn{1}{|p{27.3519231pt}}{\centering x} & \multicolumn{1}{|p{27.3519231pt}}{\centering x} & \multicolumn{1}{|p{27.3519231pt}}{\centering x} & \multicolumn{1}{|p{27.3519231pt}}{\centering x} & \multicolumn{1}{|p{27.3519231pt}}{\centering x} & \multicolumn{1}{|p{27.3519231pt}}{\centering } & \multicolumn{1}{|p{27.3519231pt}}{\centering } & \multicolumn{1}{|p{27.3519231pt}}{\centering x} & \multicolumn{1}{|p{27.3519231pt}|}{\centering x}\\ 
\hline
\multicolumn{1}{|p{100.33pt}}{\raggedright User Relationship (10)} & \multicolumn{1}{|p{100.44pt}}{\raggedright User Relationship data} & \multicolumn{1}{|p{27.3519231pt}}{\centering x} & \multicolumn{1}{|p{27.3519231pt}}{\centering } & \multicolumn{1}{|p{27.3519231pt}}{\centering } &\multicolumn{1}{|p{100.55pt}}{\raggedright Find friends via contact list (6)} & \multicolumn{1}{|p{27.3519231pt}}{\centering x} & \multicolumn{1}{|p{27.3519231pt}}{\centering x} & \multicolumn{1}{|p{27.3519231pt}}{\centering x} & \multicolumn{1}{|p{27.3519231pt}}{\centering x} & \multicolumn{1}{|p{27.3519231pt}}{\centering x} & \multicolumn{1}{|p{27.3519231pt}}{\centering } & \multicolumn{1}{|p{27.3519231pt}}{\centering x} & \multicolumn{1}{|p{27.3519231pt}}{\centering } & \multicolumn{1}{|p{27.3519231pt}|}{\centering x}\\ 
\hline 
\end{longtable}
\end{sidewaystable*}   

\twocolumn

\noindent\textbf{Other system features. }
The Messaging and Moments functions are similar to other social applications. WeChat does not directly process and analyze the messages and moments sent by users. Therefore, when the above behaviors leak other people's IDP information, there are no effective measures for awareness, authorization, and access, but some efforts have been made in accountability, auditability, and alignment. First of all, WeChat has opened a reporting channel. When users find that their private information has been leaked due to other users' behavior, they can file a complaint with WeChat officials. WeChat guarantees that it will handle the report according to the law and its own privacy policy, but the basis and efficiency of the handling are not yet clear.

In our analysis of WeChat, we noted variable effectiveness regarding IDP risk mitigation across system features. On one hand, a major vulnerability exists in WeChat Pay, where users can view partial recipient information without cost, posing a significant privacy risk due to inadequate restrictions or encryption measures. On the other hand, WeChat successfully manages the privacy of contacts by i) encrypting contact information upon upload and using it only in encrypted form and solely for matching purposes, and ii) informing users properly about this process; this effectively mitigates a potential IDP risk. 
On the contrary, for content shared on public channels, WeChat conducts manual reviews and responds to user complaints to address privacy infringements. Although this practice is standard across the industry, it's inherently reactive, which can be inefficient and allow for initial exposure of private information.

\vspace{-3mm}
\section{Conclusions}
\label{sec:conc}

In this paper, we introduced a novel threat modeling approach, IDPA, specifically targeting interdependent privacy risks, where the privacy of an individual (user or even non-user) can be harmed by other users (and other stakeholders) of a system. IDPA is inspired by LINDDUN as far as using data flows and threat categories as a starting point; however, it is explicitly designed for uncovering IDP threats through an enriched data flow diagram, IDP-specific threat categories, and misactor analysis. Furthermore, IDPA benefits from being an IDP-focused methodology; by concentrating on awareness, consent, and access control, the threat modeling process is streamlined and caters to the average practitioner. IDPA helps analysts understand and alleviate some of the pain points of privacy threat protection. Privacy protection usually requires knowing the details of the data and finding solutions based on this. However, IDPA only requires the system to know whether the data is related to the authorized data sender and can use this as a starting point for subsequent analysis and finding mitigation solutions. IDPA balances the functionality of the system with privacy protection. Under the premise that the law does not have clear rules for IDP, it emphasizes the analysis of the motivations of misactors and leads to mitigation measures such as providing voluntary tools and methods for users who are willing to protect the privacy of others. IDPA also provides the system with more ideas to alleviate IDP problems. Although the implementation of measures such as authorization and access control is very difficult in IDP scenarios, accountability, auditability, and alignment can better prevent and hold misactors accountable for privacy leaks. We validated our methodology by conducting a case study on the popular WeChat application, exposing various IDP issues and qualitatively assessing the efficiency of countermeasures already in place (potentially inadvertently) in terms of interdependent privacy. 

\noindent\textbf{Limitations. }
The limitations of this study come from the inherent complexity of IDP threats, which often involve multiple actors, unclear responsibilities, and a lack of visibility into data flows. The real world currently lacks sufficient precedents or reliable methods to systematically address such threats, making it challenging to design comprehensive mitigation strategies.

One key shortcoming is the incomplete articulation of the 6A principle, particularly in the area of accountability. Although IDPA introduces accountability as a core consideration, it does not yet offer a concrete mechanism for assigning responsibility in cases where users leak others' private data and the system cannot assess the data content. This raises unresolved questions: How should a system, acting merely as an intermediary, respond to such privacy violations? What rules should it implement to ensure ethical data handling without overstepping technical or legal boundaries? 

Due to the shortage of real cases for solving IDP, we cannot give more examples as references at present. In addition, the solutions mentioned in this article cannot mitigate IDP threats completely.

\noindent\textbf{Future work. }
Our first area of future work will be to design specific accountability mechanisms within the 6A framework, particularly for situations where the system does not have full visibility into the data being shared. This includes designing rules that guide platform responses when users leak others' data and redefining privacy policies to support proactive protection as opposed to risk avoidance. We also plan to explore normative models and policy structures that help clarify responsibility in multi-actor IDP scenarios, addressing current gaps caused by the absence of real-world precedents.

Although we motivated and validated IDPA via the third-party app ecosystem and WeChat, we foresee that IDPA generalizes well to a broad range of software systems, much like LINDDUN. As far as future work, we aim to better embed IDPA into theoretical/conceptual privacy frameworks and conduct case studies on multiple third-party apps and platforms, and other networked software systems.

We are also eager to investigate whether threat modeling can benefit from specialized approaches across privacy threat ``types'' besides IDP. The emergence of LINDDUN MAESTRO\footnote{\url{https://linddun.org/instructions-for-maestro/}}, a more elaborate LINDDUN version with enriched system models and multiple viewpoints catering for specific threat types, points towards this direction. 

\appendix

\section{Notation}

\begin{table}[!h]
\centering
\caption{List of Abbreviations}
\label{tab:abbreviations}
\begin{tabular}{@{}ll@{}}
\toprule
\textbf{Abbreviation} & \textbf{Meaning}                       \\ \midrule
PTM                   & Privacy threat modeling                \\
IDP                   & Interdependent privacy                 \\
IDPA                  & Interdependent privacy analysis        \\
IDPF                  & Privacy-Interdependent Data Flow        \\
PIDPF                  & Potentially Privacy-Interdependent Data Flow \\
NIDPF                  & Normal data flow                       \\
MU                    & Malicious users                        \\
IU                    & Indifferent users                      \\
UU                    & Unprepared users                       \\
UFU                   & Uninformed users                       \\
SP                    & Service providers                      \\
GA                    & Government authorities                 \\
IS                    & Improper sharing of privacy-interdependent data \\
IST                   & Improper storage of privacy-interdependent data \\
IP                    & Improper processing of privacy-interdependent data \\
\bottomrule
\end{tabular}
\end{table}

\end{document}